\documentclass[fleqn,usenatbib]{mnras}
\usepackage{newtxtext,newtxmath}
\usepackage[T1]{fontenc}

\DeclareRobustCommand{\VAN}[3]{#2}
\let\VANthebibliography\thebibliography
\def\thebibliography{\DeclareRobustCommand{\VAN}[3]{##3}\VANthebibliography}

\usepackage{graphicx}
\usepackage{amsmath}
\usepackage{bm}
\usepackage{xr-hyper}

\makeatletter
\newcommand*{\addFileDependency}[1]{%
  \typeout{(#1)}%
  \@addtofilelist{#1}%
  \IfFileExists{#1}{}{\typeout{No file #1.}}%
}
\makeatother

\newcommand*{\myexternaldocument}[1]{%
    \externaldocument{#1}%
    \addFileDependency{#1.tex}%
    \addFileDependency{#1.aux}%
}

\myexternaldocument{appendices/appendices}

\newcommand{\appropto}{\mathrel{\vcenter{
  \offinterlineskip\halign{\hfil$##$\cr
    \propto\cr\noalign{\kern2pt}\sim\cr\noalign{\kern-2pt}}}}}

\newcommand{\sgn}{\text{sgn}}

\defcitealias{NFW96}{NFW}
\defcitealias{Oman+19}{O19}
\defcitealias{SantosSantos+20}{SS20}

\usepackage{hyperref}

\title[Rotation curves, cusps and cores]{The diversity of rotation curves of simulated galaxies with cusps and cores}

\author[F. A. Roper et al.]{\newauthor
Finn A. Roper$^{1,2,3}$\thanks{E-mail: finn.roper@ed.ac.uk,  kyle.a.oman@durham.ac.uk},
Kyle A. Oman$^{2,3}$\footnotemark[1]{},
Carlos S. Frenk$^{2,3}$,
Alejandro Ben{\'i}tez-Llambay$^{4}$,
Julio F. Navarro$^{5}$,\newauthor
and Isabel M.E. Santos-Santos$^{2,3,5}$
\\
$^{1}$ Institute for Astronomy, University of Edinburgh, Royal Observatory, Blackford Hill, Edinburgh EH9 3HJ, United Kingdom\\
$^{2}$ Institute for Computational Cosmology, Durham University, South Road, Durham DH1 3LE, United Kingdom\\
$^{3}$ Department of Physics, Durham University, South Road, Durham DH1 3LE, United Kingdom\\
$^{4}$ University of Milano-Bicocca, Piazza della Scienza, 3, 20126 Milano MI, Italy\\
$^{5}$ Department of Physics and Astronomy, University of Victoria, Victoria, BC, Canada V8P 5C2\\
}

\date{Accepted XXX. Received YYY; in original form ZZZ}

\pubyear{\the\year}

\begin{document}
\label{firstpage}
\pagerange{\pageref{firstpage}--\pageref{lastpage}}
\maketitle

\begin{abstract}
  We use $\Lambda$CDM cosmological hydrodynamical simulations to explore the kinematics of gaseous discs in late-type dwarf galaxies. We create high-resolution 21-cm `observations' of simulated dwarfs produced in two variations of the EAGLE galaxy formation model: one where supernova-driven gas flows redistribute dark matter and form constant-density central `cores', and another where the central `cusps' survive intact. We `observe' each galaxy along multiple sight lines and derive a rotation curve for each observation using a conventional tilted-ring approach to model the gas kinematics. We find that the modelling process introduces systematic discrepancies between the recovered rotation curve and the actual circular velocity curve driven primarily by (i) non-circular gas orbits within the discs; (ii) the finite thickness of gaseous discs, which leads to overlap of different radii in projection; and (iii) departures from dynamical equilibrium. Dwarfs with dark matter cusps often appear to have a core, whilst the inverse error is less common. These effects naturally reproduce an observed trend which other models struggle to explain: late-type dwarfs with more steeply-rising rotation curves appear to be dark matter-dominated in the inner regions, whereas the opposite seems to hold in galaxies with core-like rotation curves. We conclude that if similar effects affect the rotation curves of observed dwarfs, a late-type dwarf population in which all galaxies have sizeable dark matter cores is most likely incompatible with current measurements.
\end{abstract}


\section{Introduction} \label{Sec:Introduction}

The cusp-core problem has been an enduring one in cosmology, and represents an important challenge to our current understanding of dark matter (DM) and galaxy formation in a $\Lambda$CDM universe. The structure of cold DM halos has be studied extensively using cosmological N-body simulations, and is now well understood (\citealt*{NFW96,Navarro+97}; \citealt{Wang+20}). Such simulations typically produce haloes with density profiles rising to an asymptotic `cusp' in the centres of all haloes expected to host galaxies. In contrast, many observed dwarf galaxies have rotation curves implying a constant central `core' in DM density \citep[][and see \citealp{deBlok10} for a review]{Flores&Primack94,Moore94}. A resolution of this apparent overprediction of the central DM density in dwarfs remains elusive.

Many possible explanations for this conflict have been proposed. \citet{Oman+15} argued that in the context of a DM cosmology, the possibilities fall into three categories: (i) the DM differs from the fiducial hypothesis of cold, collisionless particles; (ii) DM cusps are transformed into cores, e.g. through gravitational coupling to violent gas motions driven by supernova (SN) explosions; (iii) the DM distributions within galaxies have been incorrectly inferred from kinematic data. Of course, combinations of effects in more than one of these categories are also possible.

It has been shown that allowing for departures from the fundamental assumption in $\Lambda$CDM cosmology that DM is a cold, collisionless fluid can affect the internal kinematics of dwarf galaxies. Self-interacting dark matter (SIDM) models have been proposed \citep{Spergel&Steinhardt00}, where scattering interactions between the cold DM particles are introduced. Interaction cross-sections per unit mass of ${\sigma/m \sim 1 \, \mathrm{cm}^{2} \, \mathrm{g}^{-1}}$ lead to the thermalisation of the centres of DM haloes needed to produce cores in dwarfs, whilst leaving the successes of CDM on larger scales essentially intact \citep[see e.g.][]{Rocha+13,Elbert+15,Ren+19}. It has been argued that SIDM models may explain the broad diversity in dwarf rotation curve shapes, although additional effects due to galaxy formation physics complicate this picture \citep{Creasey+17,Kaplinghat+20}.

DM cores may also be created through `baryonic' processes. \citet*{NEF96} showed that a sudden loss of mass from the inner halo, which may occur when gas is expelled by supernova explosions, can redistribute DM away from the halo centre, creating a core. This process is reversible as baryons may reaccrete over time and recreate a cusp. However, if these blowouts are sufficiently frequent and numerous, the end result may be a shallower central density profile \citep[][]{Pontzen&Governato12}. For significant baryon-induced core creation (BICC) to occur, the baryons must be sufficiently gravitationally dominant to cause the halo to contract before the subsequent expulsion of gas from the centre \citep{BenitezLlambay+19}. These results suggest that DM cores may form in dwarf galaxies with sufficiently variable and energetic central star formation (SF) feedback \citep[][]{Faucher-Giguere18}, and this has been seen in multiple cosmological galaxy formation simulations \citep[e.g.][]{DiCintio+14,Chan+15,Tollet+16,Verbeke+17,Hopkins+18,BenitezLlambay+19,Jahn+21}.

In principle, `beam smearing' \citep[see][and references therein]{Swaters+09} or ambiguity in the mass-to-light ratios of galaxies can lead to an incorrect inference of the DM distribution, giving the appearance of a core where none exists. However, cores are still inferred to exist from high-resolution 21-cm radio observations of dwarf galaxies, such as those taken for the THINGS \citep{Walter+08} and LITTLE~THINGS \citep{Hunter+12} surveys, where concerns around beam smearing are greatly alleviated. The mass-to-light ratios of galaxies in these surveys were derived from an empirical relation based on the optical colours of galaxies \citep{Bell&deJong01,Bruzal&Charlot03}, which has been robustly validated using multiple techniques \citep[see e.g.][]{Oh+11,Oh+15}.

Nevertheless, the possibility that observed cores are a mere symptom of systematic issues in observation or modelling has not been definitively ruled out. The shapes of DM haloes have been predicted to be triaxial in shape since some of the earliest CDM simulations \citep{Davis+85,Frenk+88}. This generates an aspherical gravitational potential, with the central regions having the greatest asphericity \citep*{Hayashi+07}. A disc galaxy at the centre of such a halo, typically aligned on a plane close to that defined by the intermediate and major axes of the halo \citep{HayashiAndNavarro06}, therefore inhabits a non-axisymmetric potential. Non-circular motions (NCMs) in the disc are induced as gas orbits are elongated in the potential, manifesting as mainly bisymmetric (${m = 2}$ harmonic) fluctuations in azimuthal velocity as a function of projection angle \citep[][hereafter \citetalias{Oman+19}; see also \citealt{Marasco+18}]{Oman+19}. Even very slight asphericity in the potential can result in significant NCMs in the disc \citep[e.g.][]{HayashiAndNavarro06}. Although harmonic decomposition of the velocity fields of observed dwarfs suggest that the amplitudes of NCMs are relatively small \citep{Trachternach+08}, these could easily have been underestimated during model fitting. Often, variations in velocity associated with NCMs are absorbed by other model parameters that are degenerate with NCMs (see Sec.~\ref{subsec:ncms}). This may lead to surprisingly large errors in the measurement of rotation curves, or, in other words, cause large discrepancies between the inferred and true circular velocity at the same radius \citepalias{Oman+19}.

The weak, bisymmetric, bar-like velocity patterns discussed here are distinct from the far stronger patterns characteristic of barred spiral galaxies \citep[e.g.][]{Machado&Athanassoula10}. Those patterns form through an instability occurring in massive, self-gravitating discs \citep{Ostriker&Peebles73,Algorry+17}, but the weaker bar-like perturbations we are concerned with are instead suppressed in such systems \citep[see][]{Marasco+18} because massive discs `sphericalise' their initially triaxial host haloes as they form \citep[see][and references therein]{Abadi+10}. In DM-dominated dwarf galaxies, this process is less effective and so significant velocity perturbations (and therefore NCMs) due to halo triaxiality can persist.

\citetalias{Oman+19} showed that NCMs may affect the rotation speed inferred at any given radius, depending on the orientation of the projection axis relative to the main kinematic axes of a gaseous disc. They also showed that conventional kinematic models systematically underestimate the circular velocity in the inner regions of galaxies, due to their inability to account for the geometric thickness of the gas discs. These effects can, and likely do, lead to many measured rotation curves rising more slowly than they should, leading to systematic underestimation of the central DM density. When such systematic effects are accounted for, the diversity of inferred rotation curve shapes of simulated $\Lambda$CDM galaxies may be reconciled with that observed (although this depends on the galaxy formation model assumed).

\citet[][hereafter \citetalias{SantosSantos+20}]{SantosSantos+20} investigated multiple explanations for the observed diversity of rotation curve shapes using cosmological simulations: BICC, SIDM, and NCMs, as well as its connection with the mass discrepancy-acceleration relation \citep*[MDAR;][]{McGaugh04,McGaugh+16}. Although the presence of cores created by BICC or SIDM generated some diversity in the shapes of dwarf rotation curves, models in both categories failed to span the full range observed. They found that these explanations for the diversity in rotation curve shapes could not account for the observation that dwarf galaxies with more steeply-rising rotation curves typically have higher central DM fractions, and vice versa. Systematic errors caused by NCMs, however, were able to account for this, and apparently also for much of the observed diversity in rotation curve shapes, albeit based on an analysis of only two simulated galaxies with DM cusps.

In this work, we investigate whether DM haloes with cusps or cores (or a mix of both) are compatible with the observed diversity both in the shapes of dwarf galaxy rotation curves and the inferred baryonic matter fraction in galaxies' inner regions, once plausible systematic effects in the kinematic modelling process have been accounted for. We use selections of simulated galaxies realised with two galaxy formation models (Secs.~\ref{subsec:Simulation_selection}--\ref{subsec:Gal_selection}) -- one whose SF prescription causes the creation of cores in DM haloes hosting dwarfs, the other whose prescription preserves their initial cusps. We observe these galaxies multiple times each along independent sight lines (Secs.~\ref{subsec:Mock_observation}--\ref{subsec:Model_fitting}). These galaxy selections are both shown to produce realistic dwarf galaxy populations as judged from well-known scaling relations (Sec.~\ref{subsec:scaling_relations}) and being kinematically and geometrically similar to observed dwarfs (Secs.~\ref{subsec:ncms}--\ref{subsec:disps}).  We quantify the discrepancy between their circular velocity curves and their `observed' rotation curves, and compare the characteristics of the two `observed' populations (Sec.~\ref{sec:Results}). We discuss the physical interpretation, caveats, and implications of our results in Sec.~\ref{sec:Discussion}, and summarise in Sec.~\ref{sec:Conclusion}. Throughout this paper, we assume a cosmology with ${\mathrm{H}_0 = 70.4\,\mathrm{km}\,\mathrm{s}^{-1}\,\mathrm{Mpc}^{-1}}$ and all other cosmological parameters consistent with WMAP-7 values \citep{Jarosik+11}.

\section{Method} \label{sec:Method}

\subsection{Simulations} \label{subsec:Simulation_selection}

The simulations in this work were carried out using a modified version of the N-Body, `pressure-entropy' \citep{Hopkins13} smoothed particle hydrodynamics code {\sc Gadget-3}, itself a modified version of the {\sc Gadget-2} code \citep{Springel05}. The galaxy formation model is that of the Evolution and Assembly of GaLaxies and their Environments (EAGLE) project \citep{Schaye+15,Crain+15}. The EAGLE model includes `subgrid' prescriptions for radiative cooling, star formation, stellar mass loss, feedback and metal enrichment of surrounding gas, supermassive black hole (SMBH) gas accretion and mergers, and active galactic nuclei (AGN) feedback. The model's parameters were calibrated against present-day observations of the galaxy stellar mass function, the galaxy size distribution, and the amplitude of the SMBH-stellar mass relation. The simulations' initial conditions are set at ${z=127}$ \citep[see][appendix B, for details]{Schaye+15}) and are analysed at ${z=0}$.

\begin{figure}
\centering
\includegraphics[width=\columnwidth]{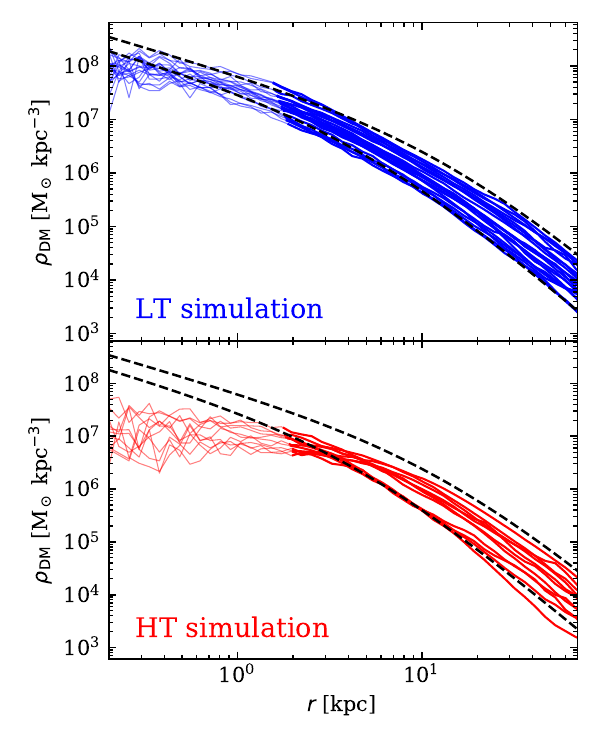}
\caption{\emph{Upper panel:} The spherically averaged DM density profiles of the $21$ dwarf galaxies selected for kinematics analysis from the LT simulation are shown with blue solid lines. The lines are thinner inside the `convergence radius' \citep[][eq.~20, with $t_\mathrm{relax}(r)\gtrsim 0.6 t_0$]{Power+03}, outside which the circular velocity curve converges to better than $\sim 10$~per~cent. Two \citetalias{NFW96} density profiles, with ${V_\mathrm{c,max}=50}$ and ${120\,\mathrm{km}\,\mathrm{s}^{-1}}$, and the median concentration given their mass \citep{Ludlow+16} are shown with dashed black lines. \emph{Lower panel:} As upper panel but for the $11$ dwarfs selected from the HT simulation, shown in red. LT dwarf haloes retain their ${\rho\sim r^{-1}}$ cusps and therefore closely follow the \citetalias{NFW96} profile shape at all resolved radii. HT dwarf haloes, however, develop ${\rho\sim r^0}$ cores with typical sizes of ${\sim 2}$--${3\,\mathrm{kpc}}$.}
\label{fig:dens_curves}
\end{figure}

We use two variations based on the `Recal' calibration of the EAGLE model, originally presented by \citet{BenitezLlambay+19}. The main difference between the two is the value chosen for the parameter controlling the hydrogen number density threshold above which SF is allowed to occur, $n_\mathrm{thr}$. The `low threshold' variant, which we label LT, fixes ${n_\mathrm{thr}=0.1\,\mathrm{cm}^{-3}}$. This value is close to typical values of the metallicity-dependent threshold proposed by \citep{Schaye04} used in the fiducial Recal model \citep[see][eq.~4]{Schaye+10}. The second, `high threshold' (HT) model increases the threshold to a constant ${n_\mathrm{thr} = 10 \, \mathrm{cm}^{-3}}$. The increased threshold allows gas to accumulate in the centres of dwarf galaxies, often becoming the locally dominant contributor to the gravitational force, before SF begins and subsequent SN feedback then expels substantial amounts of gas from their central regions. Gas can then reaccrete and the cycle repeats, with the bursty SF driving much greater fluctuations in the central gravitational force than occur in the LT variant where more continuous SF does not allow such significant gas accumulation and subsequent violent blowouts. In dwarf galaxies in the HT variant, the mechanism identified by \citet[][see also \citealp{Read&Gilmore05,Pontzen&Governato12}]{NEF96} operates efficiently, triggering the formation of DM cores \citep{BenitezLlambay+19}. In the LT variant, all dwarf galaxies retain their DM cusps.

We show the DM density profiles of all dwarf galaxies selected for kinematic analysis (see Sec.~\ref{subsec:Gal_selection} below) from the LT and HT simulations in the upper and lower panels of Fig.~\ref{fig:dens_curves}, respectively. The DM density profiles of LT dwarfs have shapes closely resembling the \citet*[][hereafter \citetalias{NFW96}]{NFW96} profile \citep[assuming a typical concentration;][]{Ludlow+16} at all well-resolved \citep{Power+03} radii: the dashed black lines show \citetalias{NFW96} profiles with ${V_\mathrm{c,max}=50}$~and~${120\,\mathrm{km}\,\mathrm{s}^{-1}}$. The density profiles of HT dwarfs, on the other hand, diverge from the \citetalias{NFW96} profile shape, becoming nearly flat in the inner few kiloparsecs. We note: (i) that the density profiles of HT galaxies begin to flatten well outside their \citet{Power+03} `convergence radii', inside which the lines are drawn thinner; (ii) that this convergence criterion is rather conservative as it is defined in terms of a cumulative quantity ($V_\mathrm{circ}$), but density is a differential quantity; and (iii) that the difference with respect to the LT dwarfs, which form $\sim200\,\mathrm{pc}$ cores due to numerical effects, is very clear. We are therefore confident that the cores in the HT dwarfs are not numerical artefacts.

Note that both $n_{\mathrm{thr}} = 0.1$~and~$10 \, \mathrm{cm}^{-3}$ are far below the densities at which stars realistically form \citep[$n_{\mathrm{thr}} \sim 10^{4} \, \mathrm{cm}^{-3}$;][]{Dutton+19}. Increasing $n_{\mathrm{thr}}$ is used in this work as a straightforward means of ensuring that gas accumulated in haloes becomes self-gravitating before SF occurs and so induces core formation whilst minimally altering other observables. In other simulations where gas physics is modelled differently \citep[e.g.][]{Jahn+21}, it has been shown that a high $n_{\mathrm{thr}}$ is not a necessary condition for baryon-induced cores to form, and others suggest that it may not be a sufficient condition either \citep[e.g.][]{BenitezLlambay+19,Dutton+20}. As such, it is primarily the resulting BICC, and not the value of $n_{\mathrm{thr}}$ itself, that we are concerned with in this work.

The only other difference between the two model variants is that in the HT model AGN feedback is disabled, while it is enabled in the LT model. However, AGN feedback has negligible influence on EAGLE galaxies in the range of masses on which we focus in this work \citep[${M_{\mathrm{vir}} \lesssim 10^{11.5} \, \mathrm{M}_{\sun}}$;][]{Crain+15,Bower+16} -- in fact, no SMBHs are seeded in haloes of ${M_{\mathrm{vir}} < 10^{10} \, \mathrm{M}_{\sun}} \, h^{-1}$ \citep{Crain+15}.

Both simulations (LT and HT) use identical initial conditions for a periodic cube with a side length of ${12 \, \mathrm{Mpc}}$ (comoving). The DM resolution elements have masses of ${m_{\mathrm{DM}} = 3.2 \times 10^{6} \, \mathrm{M}_{\sun}}$, and the gas particles have initial masses of ${m_{\mathrm{gas}} = 5.3 \times 10^{5} \, \mathrm{M}_{\sun}}$, similar to the values used in the fiducial EAGLE-Recal simulations \citep{Schaye+15}. Simulated galaxies are identified using a two-step process. First, a friends-of-friends (FoF) algorithm \citep{Davis+85} identifies spatially coherent structures, with baryonic particles assigned to the FoF group of their nearest DM particle. Each FoF group is then analysed independently to search for gravitationally self-bound objects within them, using the {\sc subfind} algorithm \citep{Springel+01,Dolag+09}. These self-bound regions are referred to as `subhaloes' and typically each contains a single galaxy. The subhalo containing the particle at minimum gravitational potential within each FoF group is defined as being the `central' subhalo, with the remainder being labelled `satellite' subhaloes.

\subsection{Galaxy samples} \label{subsec:Gal_selection}

\subsubsection{Observational comparison sample}
\label{subsubsec:obs_sample}

For comparison with our sample of simulated galaxies, we select a subset of the observed galaxies compiled in \citetalias[][table~A1]{SantosSantos+20}, originally from the THINGS \citep{Walter+08,deBlok+08} and LITTLE~THINGS \citep{Hunter+12,Oh+15} surveys, and the SPARC compilation \citep*{Lelli+16}, retaining only those with ${i > 30 \degr}$, and ${R_{\mathrm{last}} > 2R_{\mathrm{fid}}}$ (see Eq.~\ref{eq:fiducial_things} for definition of $R_{\mathrm{fid}}$; $R_{\mathrm{last}}$ is defined in Sec.~\ref{subsec:Mock_observation}). We refer to the subset of this sample with ${50 < V_\mathrm{\phi,max} / \mathrm{km\ s^{-1}} < 120}$ as the `selected' observational sample throughout this work, whereas those outside this interval in $V_\mathrm{\phi,max}$ are labelled `unselected'. In some cases, our analysis requires the observed data cubes, or at least the derived velocity maps. Such data are not available for all of the galaxies in the full comparison sample above. In such cases, we use a subset further restricted to the THINGS and LITTLE~THINGS galaxies for which they are available \citepalias[see also][table~A2]{Oman+19}. 

We stress that these comparison samples are only loosely matched to our sample of simulated galaxies; in particular, the selections of observed galaxies are highly heterogeneous in the sense that they are assembled from multiple surveys and observations of individual objects, each with their own criteria for which galaxies should be observed, whereas the simulated sample is drawn from a well-defined volume.

\subsubsection{Selection of simulated galaxy samples} \label{subsubsec:sim_samples}

We select galaxies from the LT and HT simulations to obtain a sample analogous to observed late-type dwarf galaxies. We also exclude some systems not well suited to kinematic modelling, e.g. ongoing or recent mergers (see Appendix~B for details).

Our initial selection is comprised of central (not satellite; see Sec.~\ref{subsec:Simulation_selection}) galaxies with maximum circular velocities in the range ${50 < V_\mathrm{c,max} / \mathrm{km \, s^{-1}} < 120}$ (halo masses ${2 \times 10^{10} \lesssim M_{200} / \mathrm{M}_{\sun} \lesssim 10^{11.5}}$), corresponding to dwarfs but avoiding the lowest-mass systems, which are not adequately resolved in these simulations to carry out our analysis below. We then impose a minimum \ion{H}{i} gas mass, ${M_{\mathrm{HI}} > 10^{8} \, \mathrm{M}_{\sun}}$, to restrict the sample to late-types, loosely analogous to those observationally accessible to high-resolution \ion{H}{i} imaging surveys. We find that the remaining 46 galaxies from the LT simulation and 43 from the HT simulation are isolated by at least ${100 \, \mathrm{kpc}}$ from any companion satisfying the same restrictions, minimising the influence of recent/ongoing tidal interactions.

After the creation of synthetic, or `mock' observations (Sec.~\ref{subsec:Mock_observation}) of this initial sample, we proceed to a visual inspection of the systems' \ion{H}{i} surface density and intensity-weighted mean velocity maps as seen from multiple angles, and remove those obviously unsuited to analysis with a tilted-ring model (see Sec.~\ref{subsec:Model_fitting}). Excluded here are galaxies where the gas morphology is very irregular (e.g. obvious ongoing merger, or no clear rotating disc is visible), those with very strong warps, those with strong and obvious radial gas flows within the disc, or those with a companion system that is clearly gravitationally interacting. Full details of the 25~LT and 32~HT galaxies excluded in this step are included in Appendix~B. 

Finally, after carrying out the kinematic modelling step in our analysis (Sec.~\ref{subsec:Model_fitting}), the models for a few galaxies (5~LT and 3~HT) were found to completely fail to describe their corresponding mock observations; we also omit these from further consideration. The disparity in the final sample sizes (21 from the LT and 11 from the HT simulation) is due to the lower number of HT systems with quiescently rotating discs. This in turn is due to the repeated, violent outflows of gas driven by SN feedback (and the subsequent re-assembly of the gas discs) leaving the gas frequently very obviously out of kinematic equilibrium.

\subsection{Mock observations} \label{subsec:Mock_observation}

\begin{figure*}
\begin{center}
\includegraphics[width=\textwidth]{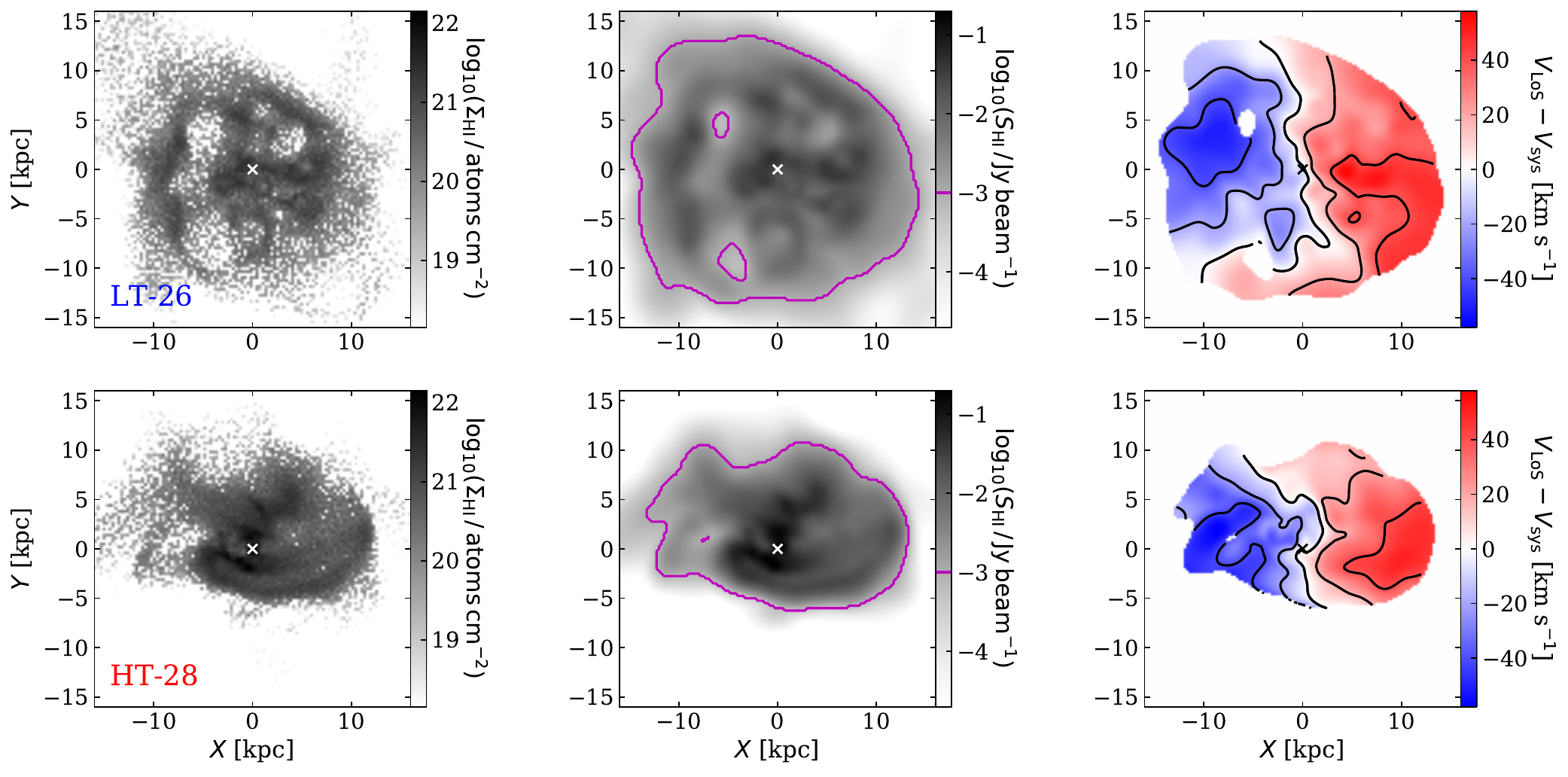}
\end{center}
\caption{Visualisations of the \ion{H}{i} gas distribution in simulated galaxies LT-26 (upper panels) and HT-28 (lower panels). \textit{Left panels:} \ion{H}{i} surface density, $\Sigma_{\mathrm{HI}}$, calculated directly from the simulation particle distribution. The spatial resolution is chosen for visual clarity. \textit{Centre panels:} 21-cm flux density $S_\mathrm{HI}$, from mock observed data cubes. \textit{Right panels:} Intensity-weighted mean velocity, from mock observed data cube. The velocity map is masked to show only pixels where the flux density exceeds $10^{-3} \, \mathrm{Jy} \, \mathrm{beam}^{-1}$ (indicated by the magenta contour in $S_\mathrm{HI}$; approximately $\Sigma_{\mathrm{HI}} = {1 \, \mathrm{M}_{\sun} \, \mathrm{pc}^{-2}} \approx 10^{20} \, \mathrm{atoms} \, \mathrm{cm}^{-2}$). Velocity contours correspond to tick locations on the colour bar. Both systems are mock observed at an inclination of $i=60\degr$ (LT-26 appears nearly face-on due to a warped outer disc, see Sec.~\ref{subsec:Mock_observation}), projection angle of $\Phi=0\degr$, position angle of $270\degr$, and assuming a distance of $12.9 \, \mathrm{Mpc}$. The mock observations are convolved with a $12 \, \mathrm{arcsec}$ ($750 \, \mathrm{pc}$) FWHM circular Gaussian beam.}
\label{fig:visualisations}
\end{figure*}

The purpose of mock observations is to provide an equivalent to telescope observations for simulated galaxies, such that the same analysis routines can be employed. Here, we mimic the characteristics of the THINGS \citep{Walter+08} and LITTLE THINGS \citep{Hunter+12} 21-cm radio surveys.

For each of our selected galaxies, we produce a set of $24$ mock observations, each corresponding to a different viewing angle. We begin by measuring the angular momentum vector of the inner $30$~per~cent (by mass) of the \ion{H}{i} disc to define a fiducial disc plane while avoiding the influence of relatively common warps in the outer discs. The $24$ sight lines are then distributed at a constant inclination of $i=60\degr$ and spaced by $\Phi=15\degr$ in azimuth\footnote{The reference azimuthal direction corresponding to $\Phi=315\degr$ is defined as $\vec{\Phi}_{315}=(1, 1, 1) - \left((1, 1, 1)\cdot\frac{\vec{L}_z}{|\vec{L}_z|}\right)\frac{\vec{L}_z}{|\vec{L}_z|}$, or $\hat{\Phi}_{315}=\frac{\vec{\Phi}_{315}}{|\vec{\Phi}_{315}|}$ as a unit vector, where $\vec{L}_z$ is the angular momentum vector defining the disc orientation. Since the galaxies are effectively randomly oriented within the simulation volume, and the vector $(1,1,1)$ (in the simulation coordinates) is arbitrarily chosen, the direction $\hat{\Phi}_{315}$, and therefore $\hat{\Phi}_{0}$, is likewise arbitrary.}. The inclination angle is within the optimal range for tilted-ring models, which struggle to capture accurately the kinematics of nearly face-on or edge-on discs. All of our mock observations have a nominal position angle (the angle clockwise from North to the approaching half of the disc) of $270\degr$. This is enforced only by aligning the component of the angular momentum vector mentioned above in the plane of the sky to point North; individual mock observations of warped or otherwise asymmetric discs may have apparent kinematic position angles differing from this value. We note that pairs of mock observations separated by $180\degr$ in $\Phi$ are not equivalent: the discs are geometrically thick and not vertically and azimuthally symmetric, which breaks this rotational symmetry. We refer to \cite[][fig. 2]{DiTeodoro&Fraternali15} for a schematic of the geometry of an observed disc.

The $24$ orientations for each galaxy are then processed with {\sc martini}\footnote{Mock Array Radio Telescope Interferometry of the Neutral ISM; \url{https://github.com/kyleaoman/martini}, version 1.5, specifically git commit identifier \verb~d1e732a~.} \citep{MARTINI}, a modular Python package for the creation of spatially resolved spectral line observations from smoothed-particle hydrodynamics simulation input \citepalias[for a detailed description, see][]{Oman+19}. This produces mock \ion{H}{i} spectroscopic `data cubes' with two spatial and one spectral axis. The mock observations are constructed assuming a distance of ${12.9 \, \mathrm{Mpc}}$ such that a $2\,\mathrm{arcsec}$ pixel corresponds to a physical size of ${125 \, \mathrm{pc}}$, and a peculiar velocity of zero. The extent of the spatial axes is the least of $128$,~$256$,~or~$512$~pixels such that the extent is larger than $R_{\mathrm{last}}$, defined as the radius of the sphere enclosing $90$~per~cent of the subhalo's \ion{H}{i} mass. ($R_{\mathrm{last}}$ for HT-17 was set to ${20\,\mathrm{kpc}}$ since gas from a companion within the same FoF group caused the radius enclosing $90$~per~cent of the \ion{H}{i} mass to be very large.) We use velocity channels with a width of ${4 \, \mathrm{km} \, \mathrm{s}^{-1}}$, and use enough channels to contain the full \ion{H}{i} velocity spectrum of each system. Each data cube is convolved along the spatial axes with a $12 \, \mathrm{arcsec}$ (FWHM) circular Gaussian beam, giving an effective physical resolution (FWHM) of approximately ${750 \, \mathrm{pc}}$. The 21-cm emission of each gas particle is calculated according to its \ion{H}{i} mass fraction. These mass fractions are corrected for self-shielding from the metagalactic ionising background following \citet{Rahmati+13}, and incorporate a pressure-dependent correction for the molecular gas fraction from \citet{Blitz&Rosolowsky06}. We label particles with \ion{H}{i} mass fractions above one~per~cent `\ion{H}{i}-bearing'. The \ion{H}{i} gas is assumed to be optically thin -- optically thick \ion{H}{i} typically occurs in compact ($\sim 100\,\mathrm{pc}$) clouds with column densities exceeding ${10^{22}\,\mathrm{atoms}\,\mathrm{cm}^{-2}}$ \citep[][see also \citealp{Allen+12}]{Braun12} which the LT and HT simulations do not resolve.

Fig.~\ref{fig:visualisations} shows example visualisations of mock observations of LT-26 and HT-28 at an inclination of $i=60\degr$, and a projection angle of ${\Phi=0\degr}$. LT-26 and HT-28 are typical of the selected LT and HT galaxy samples, respectively, and of the effect of baryon feedback on dwarfs in the samples. They are also analogous systems, with approximately equal $V_\mathrm{c,max}$ and disc mass and size (see Appendix~B) and so are chosen as illustrative examples throughout this work. The left panels show \ion{H}{i} surface density calculated on a grid by a direct summation over simulation particle masses weighted by \ion{H}{i} fraction. The centre panels show the flux density (0th moment) of the corresponding data cubes after mock observation. The right panels show the flux-weighted mean line-of-sight velocity with respect to the systemic velocity of the system (1st moment), masked to show only pixels where the flux density exceeds ${10^{-3} \, \mathrm{Jy} \, \mathrm{beam}^{-1}}$, or approximately ${\Sigma_{\mathrm{HI}} = {1 \, \mathrm{M}_{\sun} \, \mathrm{pc}^{-2}} \approx 10^{20} \, \mathrm{atoms} \, \mathrm{cm}^{-2}}$, corresponding to the typical limiting depth of observations in the THINGS and LITTLE THINGS surveys. LT-26 appears as though its inclination is less than $60\deg$. When orientated at ${\Phi=90\deg}$ however (see Appendix~E) it is clear that this is due to a warp in the outer disc. The warp is also visible in Fig.~\ref{fig:visualisations} as a twist in the outer iso-velocity contours. The central region is inclined by $60\deg$ with a position angle of $270\deg$, as intended.

\subsection{Model fitting} \label{subsec:Model_fitting}

To extract physical parameters such as rotation velocities from the data cubes, a model fitting algorithm is needed. For this, we use the 3D tilted-ring modelling tool $^{\mathrm{3D}}${\sc barolo}\footnote{\url{https://editeodoro.github.io/Bbarolo/}, version 1.5.3, specifically git commit identifier \verb~984fe8f~.} \citep{DiTeodoro&Fraternali15}, a publicly available tool designed to derive rotation curves of galaxies using emission-line observations, such as \ion{H}{i} emission. It models the disc as a number of concentric rings, each having its own set of physical parameters (inclination, rotation speed, etc.). The algorithm is `3D' in the sense that it evaluates a figure of merit for a model by comparing the actual and predicted emission in the full data cube, rather than in integrated maps of velocity, dispersion, etc. (the classical `2D' approach).

To analyse our mock observations, we use rings of $750\,\mathrm{pc}$ radial width (6 pixels, equivalent to 1 beam width, i.e. consecutive rings contain nearly independent emission), and a number of rings sufficient to enclose $R_{\mathrm{last}}$. Each ring is centered within the data cube, which by construction is the location of the potential minimum of the target galaxy. This potential minimum is generally well traced (within about ${150\,\mathrm{pc}\approx 3\,\mathrm{arcsec}}$) by the peak of the stellar light distribution \citepalias[see e.g.][fig.~2]{Oman+19}, implying that the centres that we assume for our simulated galaxies are equivalent to those which can be measured for observed galaxies. The systemic velocity is fixed to the Hubble recession velocity at a distance of ${12.9\,\mathrm{Mpc}}$ (recalling that the mocks are created with zero peculiar velocity). The \ion{H}{i} scale height of each ring, whose precise value influences the analysis very little, is set at ${100 \, \mathrm{pc}}$, following \citet[][see their sec.~7.1 for further discussion]{Iorio+17}. The \ion{H}{i} column density is fixed such that the integrated flux along the spectral axis in each pixel in the model is equal to the integrated flux in the corresponding pixel in the mock observations, i.e. $^{\mathrm{3D}}${\sc barolo}'s `local normalisation' option. This prevents strongly non-uniform regions of the disc, e.g. gas overdensities or holes, from disproportionately biasing the model \citep[][]{Lelli+12}. We have tested that adopting $^{\mathrm{3D}}${\sc barolo}'s `azimuthal normalisation' option instead produces qualitatively equivalent results \citep[see][for details of normalisation options]{DiTeodoro&Fraternali15}.

\begin{table}
\centering
\caption{Ring parameters in $^{\mathrm{3D}}${\sc barolo}. Columns from left to right: the parameters describing each ring; the symbols used for each parameter; whether the parameter is fixed or is free to be adjusted during fitting; the value at which the parameter is fixed or the bounds (if applicable) if the parameter is free; and lastly, the units used for the values given.}
\begin{tabular}{l c c r l}
\hline
Ring parameter          & Symbol              & Fixed?       & Value                     & Units \\
\hline
Centre coordinates      & $(x_{0},y_{0})$     & \checkmark   & $n_{\mathrm{px}}/2-0.5$   & $\mathrm{px}$ \\
Radial position         & $r$                 & \checkmark   & \textit{per ring}         & $\mathrm{kpc}$ \\
Radial width            & $\Delta r$          & \checkmark   & $750$                     & $\mathrm{pc}$ \\
Systemic velocity       & $V_{\mathrm{sys}}$  & \checkmark   & $907.565$                 & $\mathrm{km} \, \mathrm{s}^{-1}$ \\
Inclination angle       & $i$                 &              & $45 - 75$                 & $\mathrm{degrees}$ \\
Position angle          & $\mathrm{PA}$       &              & $250 - 290$               & $\mathrm{degrees}$ \\
Rotation velocity       & $V_{\mathrm{\phi}}$ &              & --                        & $\mathrm{km} \, \mathrm{s}^{-1}$ \\
Velocity dispersion     & $\sigma$            &              & --                        & $\mathrm{km} \, \mathrm{s}^{-1}$ \\
Gas column density      & $\Sigma_\mathrm{HI}$& \checkmark   & \textit{per pixel}        & $\mathrm{atoms}\,\mathrm{cm}^{-2}$ \\
Gas scale height        & $Z_{0}$             & \checkmark   & $1.6000$                  & $\mathrm{arcsec}$ \\
\hline
\end{tabular}
\label{tab:ring_params}
\end{table}

Each ring is fully specified by the eight parameters listed in Table~\ref{tab:ring_params}, four of which are free to vary during the fitting process: the inclination angle, position angle, rotation velocity and velocity dispersion. The inclination and position angle are constrained to remain within ${\pm 15\degr}$ and ${\pm 20\degr}$, respectively, of their `true' values ($60\degr$ and $270\degr$, respectively). We provide initial guesses of ${i=60\degr}$, ${\mathrm{PA}=270\degr}$, ${V_\mathrm{\phi}=30\,\mathrm{km}\mathrm{s}^{-1}}$, and ${\sigma=8\,\mathrm{km}\,\mathrm{s}^{-1}}$. The output of $^{\mathrm{3D}}${\sc barolo} is minimally sensitive to the initial guesses, but is sensitive to the allowed ranges for position angle and, especially, inclination \citep{DiTeodoro&Fraternali15}, motivating us to enforce these relatively narrow allowed ranges. The model optimisation proceeds in two stages. In the first stage, all four of these parameters are free to vary. Once each ring has been independently optimised in this way, the geometric parameters (inclination and position angle) are `regularised': the values as a function of radius are fit with B\'{e}zier functions. In the second stage, the geometric parameters are kept fixed to the value of their B\'{e}zier function at the corresponding ring radius and the rotation speed and velocity dispersion are re-optimised to construct the final model. Full details of the $^{\mathrm{3D}}${\sc barolo} configuration used are listed in Appendix~C.

We tested some alternative choices for $^{\mathrm{3D}}${\sc barolo}'s configuration parameters. We have checked that replacing the B\'{e}zier regularisation functions with polynomials of $0^\mathrm{th}$, $1^\mathrm{st}$, $2^\mathrm{nd}$, or $3^\mathrm{rd}$ order produces qualitatively similar results. We have also checked that allowing $^{\mathrm{3D}}${\sc barolo} to fit a radial velocity for each ring, and removing the S/N mask also give qualitatively equivalent results. We are therefore confident that our conclusions are not sensitively dependent on our chosen configuration.

\section{Realism of simulated galaxies} \label{sec:realism}

Throughout this work, simulated galaxies are mock observed and analysed to determine how increased baryon feedback and `observation' affect their inferred rotation curves. Whether our conclusions below are applicable to real galaxies depends on whether our simulated galaxies are sufficiently similar to these, such that the observation and modelling process may plausibly lead to similar consequences. We therefore compare the overall properties and structure of our sample of simulated galaxies to those observed in this section.

\subsection{Scaling relations} \label{subsec:scaling_relations} 

\begin{figure*}
\begin{center}
\includegraphics[width=\textwidth]{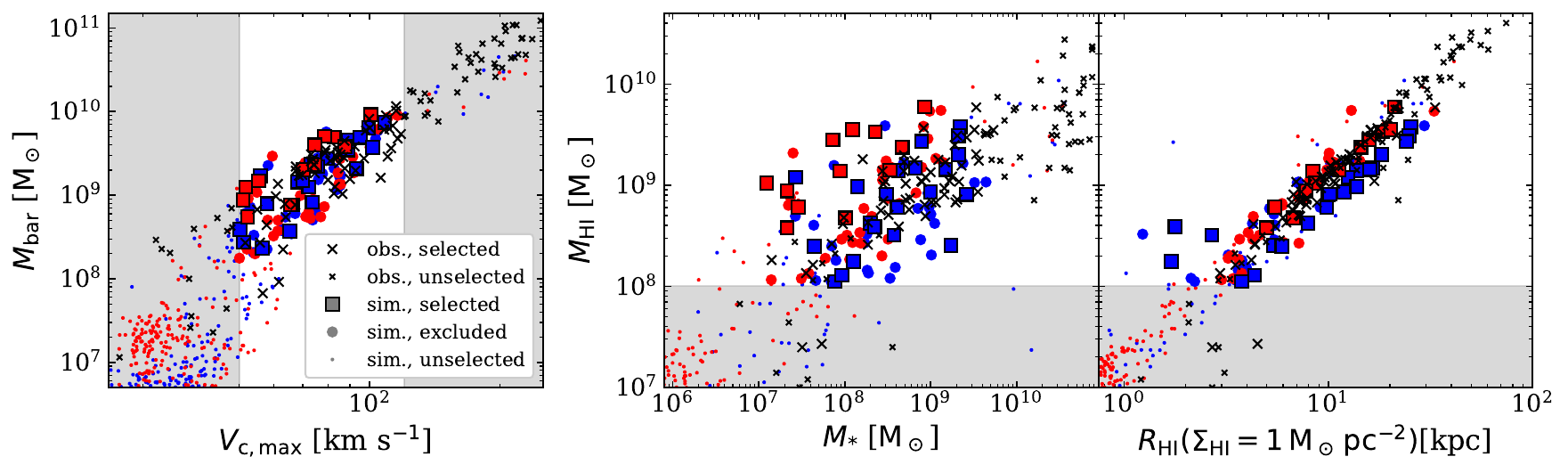}
\end{center}
\caption{\textit{Left panel:} The baryonic Tully-Fisher relation (BTFR) for all simulated galaxies in the low-SF density threshold (LT) model (blue symbols), and the high threshold (HT) model (red symbols). Galaxies with both maximum circular velocity ${50<V_\mathrm{c,max}/\mathrm{km}\,\mathrm{s}^{-1}<120}$ and total \ion{H}{i} mass ${M_\mathrm{HI} > 10^{8} \, \mathrm{M}_{\sun}}$ are indicated with larger symbols (points and squares); galaxies selected for mock observation and kinematic analysis are indicated with square symbols. The region outside of this velocity range is shaded grey. The baryonic mass is defined as ${M_{\mathrm{bar}}=M_{\star}+1.4M_{\mathrm{HI}}}$. A selection of observed galaxies from the compilation of \citetalias{SantosSantos+20} are shown with crosses, with larger symbols used for those within the same range in $V_\mathrm{\phi,max}$. \textit{Centre panel:} \ion{H}{i} mass-stellar mass relation; symbols are as in the left panel. The region corresponding to ${M_\mathrm{HI} < 10^{8} \, \mathrm{M}_{\sun}}$ is shaded grey. \textit{Right panel:} \ion{H}{i} mass-size relation; symbols and region shading are as in the centre panel. $R_\mathrm{HI}$ is defined as the galactic radius at which the \ion{H}{i} surface density falls to ${1 \, \mathrm{M}_{\sun} \, \mathrm{pc}^{-2}}$.}
\label{fig:properties}
\end{figure*}

The simulated galaxies lie on, or close to, key scaling relations, as shown in Fig.~\ref{fig:properties}. Galaxies in our final selected sample (squares) are shown in context with all central galaxies in the simulations (points), both those lying within the ranges ${50 < V_\mathrm{c,max} / \mathrm{km \, s^{-1}} < 120}$ and ${M_{\mathrm{HI}} > 10^{8}\,\mathrm{M}_{\sun}}$ but removed from our sample after visual inspection (large points; see Supplementary~Tables~B2~and~B3) and those outside these ranges (small points). These are compared to the observational comparison sample (Sec.~\ref{subsubsec:obs_sample}), including both galaxies within the same $V_\mathrm{\phi,max}$ range (large crosses) and outside (small crosses). 

The left panel of Fig.~\ref{fig:properties} shows the baryonic Tully-Fisher relation (BTFR), where our selection of galaxies from both the LT (blue symbols) and HT (red symbols) simulations align very well with the observational data. Here, the definition of $V_\mathrm{max}$ varies by dataset: for simulated systems, $V_{\mathrm{max}} = V_{\mathrm{c,max}}$ is the maximum of the circular velocity curve\footnote{In this work, `circular velocity' should be understood as this spherically-averaged quantity unless otherwise specified. In later sections we will also make use of the circular velocity determined from the gravitational acceleration in the disc midplane.}, ${V_{\mathrm{circ}}(R) = \sqrt{\mathrm{G}M(<R)/R}}$, while for observed systems $V_{\mathrm{max}} = V_{\mathrm{\phi,max}}$ is equal to the maximum velocity in their rotation curve\footnote{In this work, variables relating to circular and (observed or mock observed) rotation velocities are denoted by subscripts `c` and `$\phi$', respectively, whereas if neither is specified, the statement applies to both.}. In this figure, ${M_{\mathrm{bar}} = M_{\star} + 1.4M_{\mathrm{HI}}}$ for all systems \citep[see e.g.][]{McGaugh12}.

The centre panel shows the \ion{H}{i} mass--stellar mass relation. The stellar masses for observed galaxies assume `diet Salpeter’ and Chabrier IMFs for systems taken from the THINGS \citep{Walter+08} and LITTLE THINGS \citep{Hunter+12} surveys, respectively. For systems in the SPARC database \citep{Lelli+16}, mass-to-light ratios at $3.6\,\text{\textmu m}$ of $0.7$~and~${0.5 \, \mathrm{M_{\sun}}\,\mathrm{L_{\sun}}^{-1}}$ are assumed for bulge and disc components, respectively \citep{Lelli+16b}. For the simulated galaxies, $M_{\star}$ is a direct summation of the masses of gravitationally bound `star' particles. The simulated systems in the LT sample (blue squares) approximately follow the observed relation, whereas our selection of galaxies in the HT model (red squares) are systematically offset from both the observed relation (crosses) and the ensemble of central galaxies in the same simulation (red squares and points): they have \ion{H}{i} masses approximately ${0.5 \, \mathrm{dex}}$ higher than typical systems of the same stellar mass. The HT galaxies excluded from our selection have in many cases recently lost large amount of \ion{H}{i}, much more than their increase in stellar mass over the same period -- the same events (e.g. bursts of star formation) that triggered this loss of gas are likely responsible for their more disturbed morphologies and kinematics. Those included in our selection, on the other hand, are typically undergoing a period of rapid accretion and slow star formation. These will likely eventually undergo a burst of star formation, increasing their stellar mass and depleting their \ion{H}{i}, to bring them back towards the median relation.

The right panel shows the \ion{H}{i} mass-size relation, where size is defined as the radius at which the \ion{H}{i} surface density falls to ${\Sigma_{\mathrm{HI}} = {1 \, \mathrm{M}_{\sun} \, \mathrm{pc}^{-2}} \approx 10^{20} \, \mathrm{atoms} \, \mathrm{cm}^{-2}}$. The majority of LT systems, at fixed $M_{\mathrm{HI}}$, have approximately ${0.2 \, \mathrm{dex}}$ larger sizes than observed, while HT systems fall within the observed scatter.

\begin{figure}
\includegraphics[width=\columnwidth]{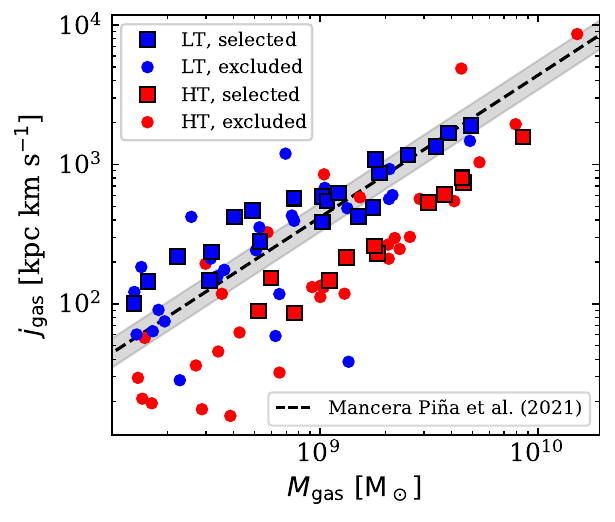}
\caption{Gas specific angular momentum, ${j_\mathrm{gas} = j_\mathrm{HI}}$, as a function of gas mass, ${M_\mathrm{gas} = 1.4M_\mathrm{HI}}$, for galaxies from the LT and HT simulations selected for kinematic analysis, shown with blue and red square markers, respectively, and excluded galaxies, shown with points of the same colours. The dashed black line shows a best-fitting {$j_\mathrm{gas}$--$M_\mathrm{gas}$} relation found by \citet{ManceraPina+21} for an expanded sample of observed galaxies relative to that used in this work: ${\log(j_\mathrm{gas}) = 1.02(\log(M_\mathrm{gas}) - 10) + 3.64}$; with an orthogonal intrinsic scatter of ${0.15\,\mathrm{dex}}$ shown by the grey band. The selected LT sample follow the observed {$j_\mathrm{gas}$--$M_\mathrm{gas}$} relation well, with slightly larger scatter, however, the selected HT sample lies at systematically higher $M_\mathrm{gas}$ than the observed relation.}
\label{fig:ang_mom}
\end{figure}

Fig.~\ref{fig:ang_mom} shows the gas mass--specific angular momentum relation, often referred to as a \citet{Fall83} relation. The empirical \mbox{$j_\mathrm{gas}$--$M_\mathrm{gas}$} relation of \citet{ManceraPina+21} based on a sample of observed dwarf galaxies from SPARC \citep{Lelli+16} and LITTLE~THINGS \citep[][see also Sec.~\ref{subsubsec:obs_sample}]{Hunter+12} as well as further late-type dwarfs from LVHIS \citep{Koribalski+18}, VLA-ANGST \citep{Ott+12}, and WHISP \citep*{vanderHulst+01} observations, is shown with the black dashed line. These observations sample the mass range ${10^{7.5} < M_{\mathrm{gas}} < 10^{10.5}}$. The selected LT sample follows the observed relation reasonably well whereas the HT sample, and particularly the subsample selected for kinematic modelling, is systematically offset to higher $M_{\mathrm{gas}}$ and/or lower $j_{\mathrm{gas}}$ values when compared to the LT sample. The offset between the LT and HT galaxies in this space is consistent with the typical differences between their \ion{H}{i} masses and sizes (see Fig.~\ref{fig:properties}). The somewhat larger \ion{H}{i} disc sizes (at fixed \ion{H}{i} mass) of LT galaxies relative to observed galaxies are compensated by their slightly lower \ion{H}{i} masses (at fixed $V_{\mathrm{c,max}}$) such that their specific angular momenta are similar to those of observed galaxies ($j_{\mathrm{gas}}\approx R_{\mathrm{HI}}V_{\mathrm{c,max}}$). Although HT galaxies have \ion{H}{i} sizes very similar to observed galaxies (at fixed \ion{H}{i} mass), their substantially higher \ion{H}{i} masses (at fixed $V_{\mathrm{c,max}}$) push them well off the observed \mbox{$j_\mathrm{gas}$--$M_\mathrm{gas}$} relation. As discussed by \citet{BenitezLlambay+19}, the large \ion{H}{i} mass of HT galaxies is due to the EAGLE feedback implementation, which becomes inefficient at pushing gas out for the HT galaxies. Furthermore, HT galaxies selected for kinematic modelling (red squares) have comparatively high average $M_\mathrm{gas}$ when compared to the full HT sample. This appears to be a selection bias, as galaxies with recent accretions of gas (following previous violent gas ejections) are more likely to be more suitable for modelling than those with recent turbulent ejections (see above discussion of the centre panel of Fig.~\ref{fig:properties}). 

Both simulated populations follow the stellar {$j_\star$--$M_\star$} relation (not shown) from \citet{ManceraPina+21}, though the selected LT and HT samples have approximately $3$ and $5$ times greater scatter than observed, respectively.

\subsection{Non-circular motions}
\label{subsec:ncms}

There is no doubt that the presence of NCMs in the gas discs of observed dwarf galaxies influences the measurement of their rotation curves (see e.g. \citealt{Valenzuela+07,Oman+15,Marasco+18}; \citetalias{Oman+19,SantosSantos+20}). It is therefore important that our simulated galaxies replicate the observed amplitudes of NCMs. Following \citet*{Schoenmakers+97}, the amplitudes of NCMs are often quantified in terms of the amplitudes, either individually or collectively, of the terms $A_{m'>0}$ in the Fourier expansion of the line-of-sight velocity, $V_{\mathrm{LoS}}$, as a function of azimuthal angle\footnote{$\phi$ refers to the angle within the plane of the disc; the angle $t$ in the plane of the sky is $t=\arctan(\tan\phi/\cos i)$.}, $\phi$, around a ring of gas:
\begin{align}
  V_\mathrm{LoS}(\phi)=V_\mathrm{sys}+\sum_{m'=1}^{\infty}a_{m'}\cos(m'\phi)+b_{m'}\sin(m'\phi),\label{eq:proj_ncm}
\end{align}
with the overall amplitude of a given order defined as ${A_{m'}=\sqrt{a_{m'}^2+b_{m'}^2}}$. We recall that a term of order $m$ in the Fourier expansion of the de-projected velocity field contributes to the ${m'=m\pm1}$ terms in Eq.~\ref{eq:proj_ncm} -- for instance, a bisymmetric pattern such as might be induced by a bar is reflected in the amplitudes $A_{1'}$ and $A_{3'}$ \citep*[e.g.][]{Schoenmakers+97}.

Multiple attempts to quantify the amplitude of NCMs from observational data have been made. \citet{Trachternach+08}, for instance, found a median radially-averaged amplitude of NCMs in THINGS galaxies of ${6.7 \, \mathrm{km} \, \mathrm{s}^{-1}}$, with approximately $90$~per~cent of galaxies with measurements of less than ${\sim 9 \, \mathrm{km} \, \mathrm{s}^{-1}}$ \citep[see also][for further discussion of a subset of the same sample; they find amplitudes ${\lesssim 10\,\mathrm{km}\,\mathrm{s}^{-1}}$ using both a similar and a complementary methodology]{Oh+11}. \citet{Oh+15} find similarly low amplitudes (${\lesssim 10\,\mathrm{km}\,\mathrm{s}^{-1}}$ in most cases, up to ${20\,\mathrm{km}\,\mathrm{s}^{-1}}$ in a few cases) for the first three harmonics in LITTLE~THINGS galaxies, but also note a crucial point: galaxy-scale NCMs are strongly degenerate with many other parameters such as the centroid, systemic velocity, position angle, and inclination (see also \citealt{Schoenmakers+97}; \citealt*{Wong+04}; \citealt{Spekkens+Sellwood07}). It is thus all too easy to minimise residuals without need for significant NCMs and so underestimate the true amplitudes of NCMs in observed galaxies. Effects associated with NCMs are then not included in the resulting model and subsequent analysis, meaning inferred rotation velocities can be profoundly in error. \citetalias{Oman+19} illustrated this further: they found that parameter degeneracies left them entirely unable to measure the amplitude of the dominant (${\sim15\,\mathrm{km}\,\mathrm{s}^{-1}}$) ${m=2}$ harmonic in two simulated dwarf galaxies from mock observations, even though these caused errors of up to ${10\,\mathrm{km}\,\mathrm{s}^{-1}}$ (${\geq 20}$~per~cent) in their rotation curves.

\begin{figure*}
\centering
\includegraphics[width=\textwidth]{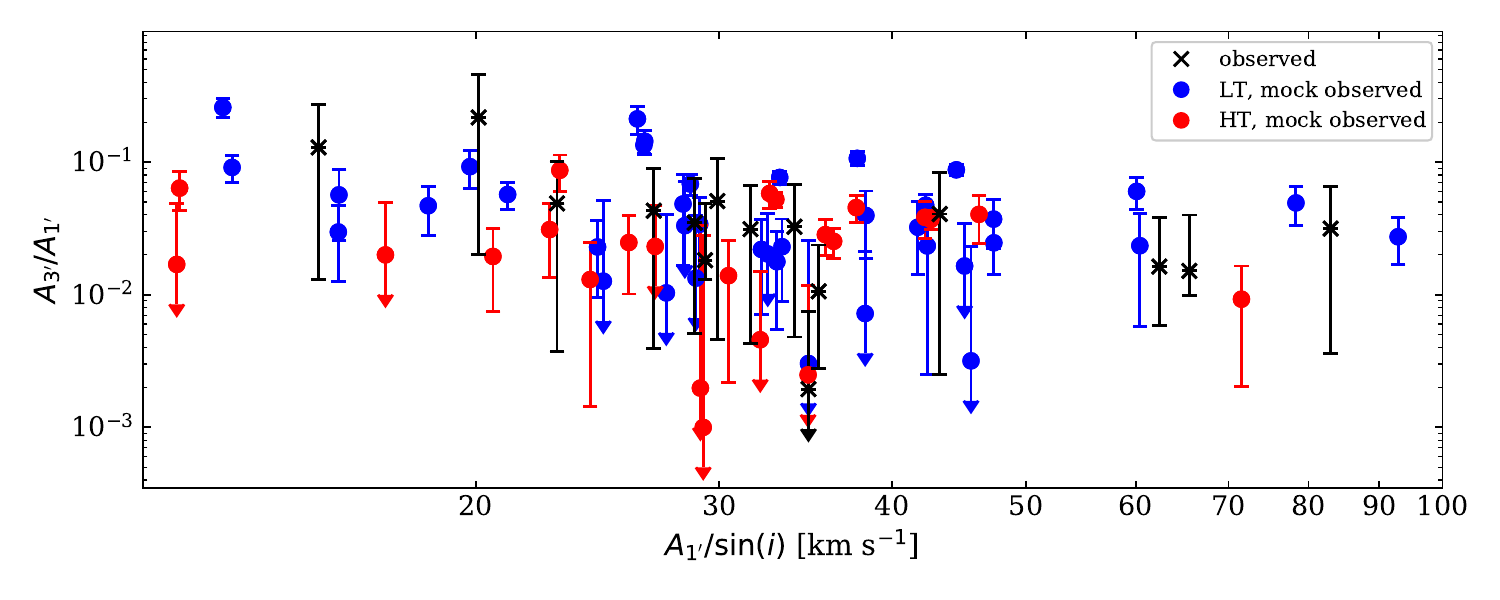}
\caption{The strength of the bisymmetric NCMs (traced by the ${m'=3}$ harmonic, see Eq.~\ref{eq:proj_ncm}) relative to the rotation speed (traced by the ${m'=1}$ harmonic), both measured at a radius of ${2\,\mathrm{kpc}}$, as a function of the inclination-corrected amplitude of the rotation speed at the same radius. Measurements for mock observed galaxies from the LT and HT simulations selected for kinematic modelling are shown with blue and red points, respectively. Measurements for selected observed galaxies from the THINGS and LITTLE~THINGS surveys, also at a radius of $2\,\mathrm{kpc}$, are shown with black crosses. We show two measurements for each simulated galaxy, for projection angles of ${\Phi=0\degr}$~and~${90\degr}$. Error bars show formal errors on $A_{\mathrm{3'}}/A_{\mathrm{1'}}$ \citep[see][for details]{Marasco+18}, and arrows indicate upper limits.}
\label{fig:ncms}
\end{figure*}

Even though the true amplitudes of a harmonic expansion of the gas velocities in a galaxy are seemingly observationally inaccessible, we may still ask: are the apparent amplitudes of NCMs seen in our simulations comparable to those in observed dwarfs? This must, however, be based on the application of an identical measurement to both observed and mock observed simulated data. This was the approach adopted by \citet{Marasco+18}: in their fig.~8, they show the relative amplitudes\footnote{$A_{m}$ in their notation is equivalent to our $A_{m'}$.} ${A_{3'}/A_{1'}}$ of the ${m'=3}$ and ${m'=1}$ harmonics as a function of $A_{1'}$ (corrected for inclination) at a given radius. The ratio of these two harmonics is indicative of the strength of the bisymmetric (${m=2}$) component of the de-projected velocity field relative to the rotation velocity at the same radius. We reproduce in Fig.~\ref{fig:ncms} their measurements at ${2\,\mathrm{kpc}}$ for THINGS and LITTLE~THINGS galaxies (we omit the SPARC sample as the required observational data cubes are typically not available) with ${50<V_\mathrm{\phi,max}/\mathrm{km}\,\mathrm{s}^{-1}<120}$. We have repeated the same measurement for our mock observations of LT and HT galaxies, shown with blue and red symbols in Fig.~\ref{fig:ncms}, for the ${\Phi=0\degr}$ and $90\degr$ orientations. Overall, the amplitude of the ${m'=3}$ harmonic, $A_{3'}$, in the simulated galaxies from both models is very similar to what is measured in observed galaxies. In the simulated galaxies, this is typically the harmonic with the largest amplitude (e.g. \citealt{Marasco+18}; \citetalias{Oman+19}). The simulated LT galaxies seem to have slightly higher $A_{3'}$ on average than both the HT and observed galaxies\footnote{Intriguingly, \citet{Jahn+21} find the opposite result: their SMUGGLE galaxy formation model, which leads to BICC, induces much stronger NCMs than a reference model where BICC does not occur. This may, however, be attributable to the presence of highly-disturbed galaxies in their sample, which would be obviously unsuitable for mass modelling, and which we have excluded from Fig.~\ref{fig:ncms} (the HT model has a larger fraction of such galaxies).}, perhaps by a factor of $2$, although the scatter in the distributions, and some cases the measurement uncertainties, are large.

\subsection{Disc thickness}
\label{subsec:thick}

\begin{figure*}
\centering
\includegraphics[width=\textwidth]{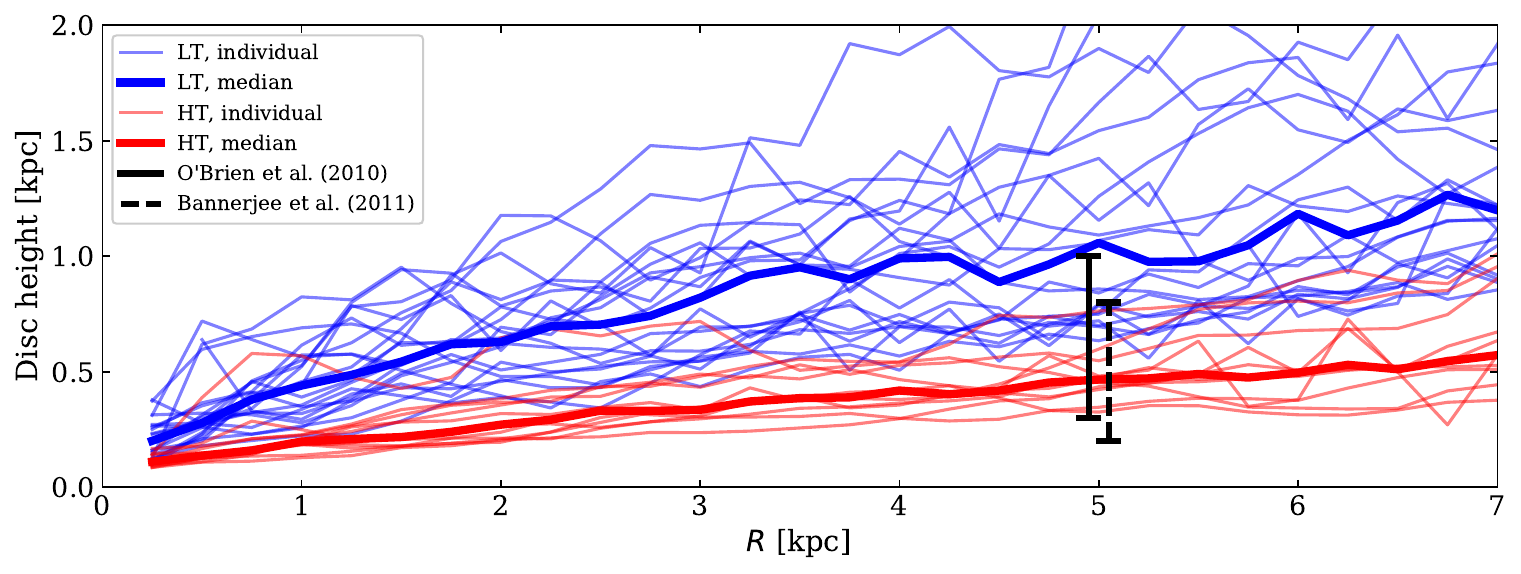}
\caption{Height above the mid-plane enclosing half of the \ion{H}{i} mass in the disc as a function of radius for the $21$ galaxies selected for kinematic analysis from the LT simulation (thin blue lines), and the $11$ selected from the HT simulation (thin red lines). The bold lines of corresponding colour show the medians of the distributions at each radius. Estimates of the half-mass disc height at ${R=5\,\mathrm{kpc}}$ of observed dwarfs from \citet{OBrien+10} and \citet{Banerjee+11} are shown as the black solid and dashed bars, respectively. The atomic gas discs of galaxies selected from the HT simulation are typically approximately half as thick as those of those selected from the LT simulation over the radial range shown, making HT dwarfs fully consistent with observation.}
\label{fig:disc_thicknesses}
\end{figure*}

Conventional kinematic models also struggle to account for the thickness of observed discs as this implies that emission from multiple rings contributes to emission along any single sight line through an inclined disc, thus introducing many degeneracies into the model (see Sec.~\ref{subsubsec:disc_disc_thickness} for further discussion). To avoid this, models typically assume thicknesses far smaller than those of observed discs which induces errors in the measured rotation curves -- i.e. if disc thicknesses were accurately known and incorporated into these models, the resulting rotation curves would be better representations of the discs' true rotation. Correspondingly, it is important that our simulated galaxies have realistic \ion{H}{i} disc thicknesses so that we obtain results comparable to analyses of observed galaxies.

Our sample of LT galaxies have very similar vertical gas structure to the APOSTLE galaxies studied by \citetalias{Oman+19}; we show their height profiles in Fig.~\ref{fig:disc_thicknesses} \citepalias[c.f.][ fig.~12]{Oman+19}. We also show the height profiles of our sample of HT galaxies in the same figure. The HT galaxies are on average a factor of ${\sim 2}$ thinner at all radii than their LT counterparts, likely due to their higher mid-plane gas densities \citep[see e.g.][]{BenitezLlambay+18}, but still are considerably thicker than the ${100\,\mathrm{pc}}$ assumed in our $^{\mathrm{3D}}${\sc barolo} configuration. Observed late-type dwarfs have estimated half-mass heights of about $0.2$--$1.0\,\mathrm{kpc}$ at a radius of $5\,\mathrm{kpc}$ (\citealp*{OBrien+10}, see their fig.~24; \citealp{Banerjee+11}; see also \citealp{Peters+17}); this range is broadly representative of further estimations of \ion{H}{i} disc heights of nearby dwarfs of similar masses \citep[see e.g.][noting that differing definitions of \ion{H}{i} height are used]{Patra20,Bacchini+22,ManceraPina+22,Li+22}. We stress that these measurements are uncertain and can require many assumptions of, e.g., vertical geometry, gravitational contributions of gas, stars, and DM, and/or hydrostatic equilibrium (see \citetalias{Oman+19}, sec.~6.2, for further discussion). \citetalias{Oman+19} established that APOSTLE dwarf galaxies (and therefore by extension the similar LT dwarfs) are likely somewhat thicker than these observations imply, by a factor of ${\lesssim 2}$. Our sample of HT dwarf galaxies therefore have gas discs with thicknesses fully consistent with these observational constraints.

\subsection{Velocity dispersion} \label{subsec:disps}

\begin{figure}
\centering
\includegraphics[width=\columnwidth]{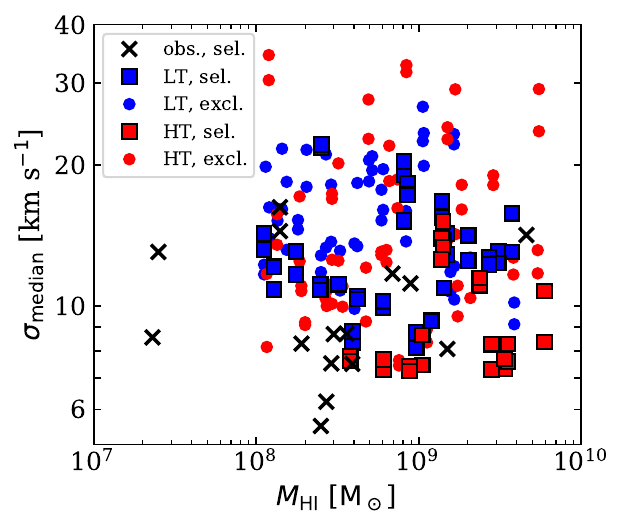}
\caption{Median line-of-sight velocity dispersion (calculated across pixels) as a function of \ion{H}{i} mass. Measurements for mock observed galaxies with ${50<V_\mathrm{c,max}/\mathrm{km}\,\mathrm{s}^{-1}<120}$ from the LT and HT simulations are shown in blue and red, respectively, with those selected for kinematic modelling shown with squares and those excluded shown with points. Measurements of selected observed galaxies from the THINGS and LITTLE~THINGS surveys are shown as black crosses. For mock observed galaxies, we show measurements for two projection angles, ${\Phi=0\degr}$~and~$90\degr$, and the median is calculated across all pixels with 21-cm flux density above ${10^{-3} \, \mathrm{Jy} \, \mathrm{beam}^{-1}}$ (corresponding approximately to \ion{H}{i} surface density above ${1 \, \mathrm{M}_{\sun} \, \mathrm{pc}^{-2}}$). For observed galaxies, it is calculated across all pixels in the S/N-masked, natural-weighted $2^\mathrm{nd}$ moment maps provided in the survey data releases. Mock observed, simulated galaxies have median velocity dispersions similar to those of the observed sample, although the LT sample has, on average, slightly higher velocity dispersions at a given \ion{H}{i} mass.}
\label{fig:disps}
\end{figure}

The thickness of an atomic gas disc is closely related to its velocity dispersion: all else being equal, a more turbulent disc is also thicker. \citetalias{Oman+19} (see their fig.~3) found that APOSTLE dwarf galaxies typically have somewhat larger velocity dispersions than observed galaxies at a given \ion{H}{i} mass. In Fig.~\ref{fig:disps}, we show that this is also true of our sample of LT galaxies selected for kinematic analysis (blue squares). The black crosses mark the measurements for the observed sample of THINGS and LITTLE~THINGS galaxies shown in Fig.~\ref{fig:ncms}. HT galaxies selected for kinematic analysis (red squares) have somewhat lower velocity dispersions than their LT counterparts, and align very well with the observed sample. When we consider instead velocity dispersion profiles (along with \ion{H}{i} surface density profiles, see Appendix~E), this is found to be true at all radii.

This likely reflects their more realistic \ion{H}{i} disc heights. We note that this observational sample contains only galaxies of sufficient kinematic regularity to be selected for mass modelling by the survey teams. Therefore, the LT and HT galaxies that we exclude from our sample for kinematic analysis (blue and red points), which often have much higher velocity dispersions, should not be compared directly with this sample (see also discussion of the centre panel of Fig.~\ref{fig:properties} in Sec.~\ref{subsec:scaling_relations}).

Observed rotation curves are often corrected for support by a radial pressure gradient\footnote{Often loosely termed an `asymmetric drift correction' by reference to the similarity of the mathematics to the analogous concept from stellar dynamics, see e.g. \citet[][appendix~A]{Valenzuela+07} for a discussion.}. This correction is estimated directly by $^{\mathrm{3D}}${\sc barolo}. In most cases the correction increases the rotation curve by only $5$~to~$10$~per~cent, essentially independently of radius, though in some rare cases it can be as large as ${\gtrsim 40}$~per~cent. These corrections thus do not play a substantive role in our results.

\section{Results} \label{sec:Results}

\subsection{Circular velocity and rotation curves of simulated galaxies} \label{subsec:circ_vel_and_rot_curves}

We show the pressure support-corrected rotation curves, $V_{\mathrm{\phi}}$, from $^{\mathrm{3D}}${\sc barolo} model fitting for each of the $24$ mock observation orientations for LT-26 and HT-28 in Fig.~\ref{fig:rot_curves} with the pale blue lines -- one curve for each galaxy is highlighted in solid magenta, corresponding to the $\Phi=0\degr$ orientation visualised in Fig.~\ref{fig:visualisations}. For this orientation, we also show the rotation curve before pressure-support correction (dotted magenta line; see Sec.~\ref{subsec:disps}). The uncertainty associated with $V_{\mathrm{\phi}}$ is taken to be the uncertainty on the rotation velocity calculated by $^{\mathrm{3D}}${\sc barolo} \citep[see][sec.~2.5]{DiTeodoro&Fraternali15}, with no contribution from the uncertainty on the pressure-support correction, as this is typically negligible. This uncertainty is shown for the $\Phi=0\degr$ orientations with the shaded magenta region\footnote{$^{\mathrm{3D}}${\sc barolo} uses an iterative algorithm to estimate errors which very occasionally fails to converge. This is the case for the outermost ring in the model for HT-28 at $\Phi=0\degr$.}.

These pressure support-corrected $V_{\mathrm{\phi}}$ curves are compared with the circular velocity curve (determined from the gravitational acceleration field in the disc midplane; black line) and the median azimuthal velocity as a function of radius of \ion{H}{i}-bearing simulation particles ($V_{\mathrm{az}}$; green line) of each system. Using the magenta curves, we illustrate our definition of an outer, `maximum' rotation velocity, $V_{\mathrm{\phi,max}}$, and an inner, `fiducial' rotation velocity, $V_{\mathrm{\phi,fid}}$ (dashed magenta lines, as labelled).

For simulated galaxies, $V_{\mathrm{\phi,max}}$ is determined as the asymptotic flat rotation velocity, or if this is ill-defined, ${V_{\mathrm{\phi,max}} \equiv V_{\mathrm{\phi}}(10 \, \mathrm{kpc})}$ (or ${V_{\mathrm{\phi}}(R_{\mathrm{last}})}$, if ${R_{\mathrm{last}} < 10 \, \mathrm{kpc}}$; see Appendix~D for details of the definition of $V_\mathrm{\phi,max}$). For circular velocity curves, the asymptotic flat rotation velocity is generally equivalent to the true maximum velocity of the curve and can be used interchangeably, but for rotation curves this is often not the case: they are more irregular, and a local upward fluctuation can often cause the maximum of the curve to overestimate the asymptotically flat velocity. Our adopted definition minimises the influence of such local fluctuations in the rotation curves.

For $V_{\mathrm{fid}}$, we follow \citetalias{SantosSantos+20} and define:
\begin{equation} 
  V_{\mathrm{fid}} = V_{\mathrm{\phi}}(R = R_{\mathrm{fid}}); \\
  \frac{R_{\mathrm{fid}}}{\mathrm{kpc}} = \frac{V_{\mathrm{\phi,max}}}{35 \, \mathrm{km} \; \mathrm{s}^{-1}}.
    \label{eq:fiducial_things}
\end{equation}
This definition adapts to the typical size of galaxies of given $V_\mathrm{max}$ to yield an inner rotation velocity tracing their central matter content. $R_{\mathrm{fid}}$ is also typically a radius that can be resolved by most high-resolution \ion{H}{i} observations of local dwarfs \citep[see e.g.][]{Hunter+12}, enabling comparison with observations.

\subsection{Rotation curve shape and central baryon dominance}
\label{subsec:rotcurshapes}

\subsubsection{Rotation curve shapes}

\begin{figure}
\begin{center}
\includegraphics[width=\columnwidth]{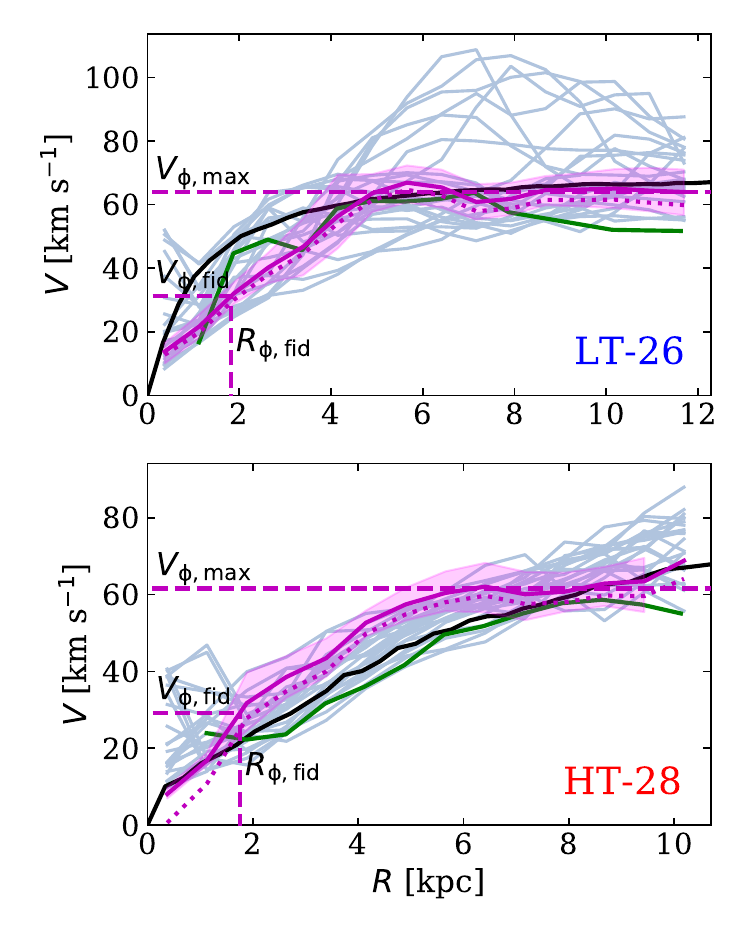}
\end{center}
\caption{Orbital velocities as a function of radius for the gas discs of simulated galaxies LT-26 and HT-28, the same example galaxies shown in Fig.~\ref{fig:visualisations}. The circular velocity, ${V_{\mathrm{circ}}(R) = \sgn(a_{r}(R))\sqrt{R|a_{r}(R)|}}$, where $a_{r}$ is the radial component of the gravitational acceleration field in the disc midplane, is shown in black. The median azimuthal velocity of \ion{H}{i}-bearing particles, $V_\mathrm{az}$, reflecting the actual motion of the gas, is shown in green. The $24$ solid blue and magenta curves show the rotation velocity, $V_{\mathrm{\phi}}$, obtained by modelling mock observations of the galaxies seen along $24$ different sight lines at fixed inclination ($60\degr$). These have been corrected for pressure support. The pressure support-corrected $V_{\mathrm{\phi}}$ from mock observations for the projection angle $\Phi=0\degr$, corresponding to the images shown in Fig.~\ref{fig:visualisations}, is shown in solid magenta, with uncertainty shown by the magenta shading. The corresponding $V_{\mathrm{\phi,max}}$, $V_{\mathrm{\phi,fid}}$, and $R_{\mathrm{\phi,fid}}$ are shown with dashed magenta lines (see Eq.~\ref{eq:fiducial_things} and Appendix~D for definitions). The rotation velocity without pressure support-correction for $\Phi=0\degr$ mock observations is shown with dotted magenta lines.}
\label{fig:rot_curves}
\end{figure}

\begin{figure*}
\begin{center}
\includegraphics[width=\textwidth]{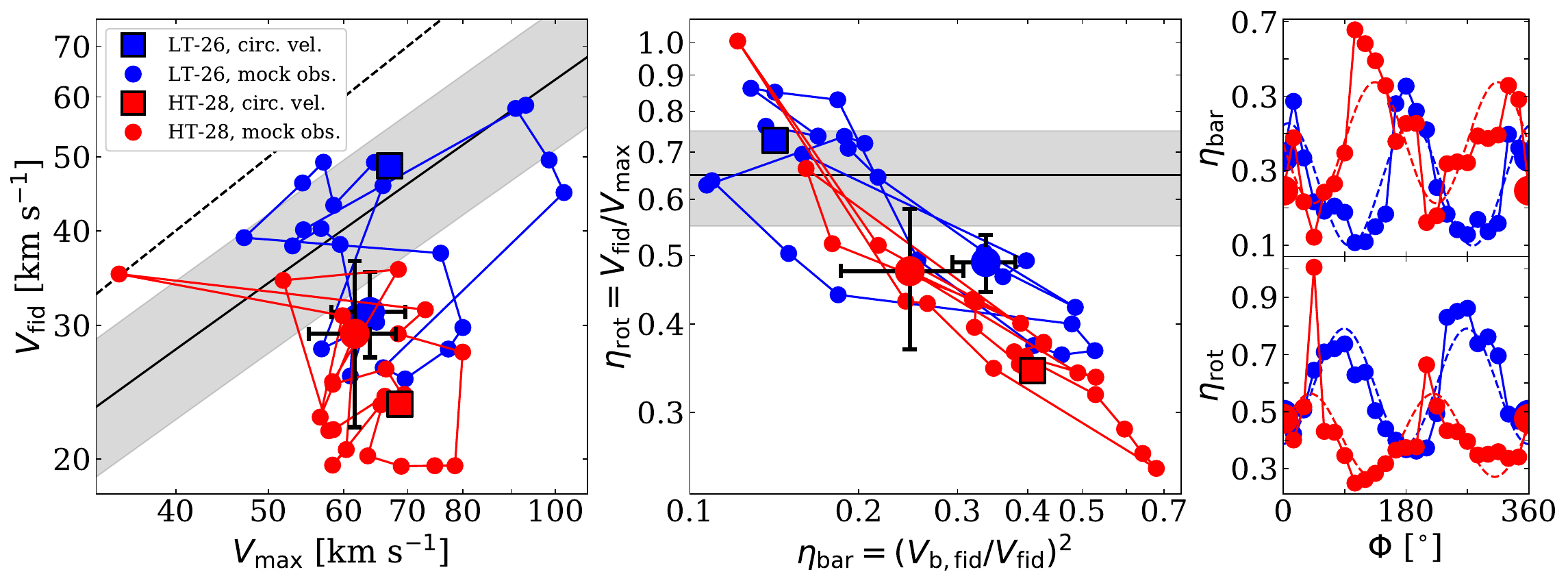}
\end{center}
\caption{\textit{Left panel:} Inner `fiducial' velocity (see Eq.~\ref{eq:fiducial_things} and Fig.~\ref{fig:rot_curves}) as a function of outer `maximum' velocity (see Appendix~D). The square markers show the values from the circular velocity curves (derived from the gravitational acceleration in the disc midplane) of galaxies LT-24 (blue) and HT-24 (red). The blue and red points show the values from each of the $24$ rotation curves obtained from mock observations for the same two galaxies. The measurement for projection angle $\Phi=0\degr$, shown in Fig.~\ref{fig:visualisations} and highlighted in Fig.~\ref{fig:rot_curves}, is shown as an enlarged point with error bars. This is connected to all measurements for values of $\Phi$ in $15\degr$ increments by solid lines. The solid black line shows the relation for an average \citetalias{NFW96} halo model assuming the mass-concentration relation of \citet{Ludlow+16}, and the shaded band illustrates the $10^\mathrm{th}$--$90^\mathrm{th}$ percentile scatter around this relation. The dashed black line shows ${V_{\mathrm{fid}} = V_{\mathrm{max}}}$. \textit{Centre panel:} Rotation curve shape parameter, $\eta_{\mathrm{rot}}$, as a function of central baryon-to-total mass fraction, $\eta_{\mathrm{bar}}$ (see Sec.~\ref{subsubsec:baryon_dom}). The solid black line shows the average $\eta_{\mathrm{rot}}$ expected of \citetalias{NFW96} cusps and the grey band shows the approximate $10^\mathrm{th}$--$90^\mathrm{th}$~percentile scatter around this value. Symbol shapes and colours, and connecting lines, are as in the left panel. \textit{Right panels:} $\eta_{\mathrm{bar}}$ and $\eta_{\mathrm{rot}}$ as functions of viewing angle, $\Phi$. Symbols and the solid lines have the same meaning as in the other panels. Dashed lines show best fitting sinusoidal functions with a $180\degr$ period, revealing the systematic effect bisymmetric NCMs have on $\eta_{\mathrm{bar}}$ and $\eta_{\mathrm{rot}}$ in both models.}
\label{fig:loops}
\end{figure*}

While a qualitative impression of the structure of the galaxies can be obtained by inspection of the curves in Figs.~\ref{fig:dens_curves}~and~\ref{fig:rot_curves}, in order to quantify the connection between the dark and baryonic matter distributions we again follow \citetalias{SantosSantos+20}. We show in the left panel of Fig.~\ref{fig:loops} (similar to their fig.~2) $V_\mathrm{fid}$ plotted against $V_\mathrm{max}$, which quantifies how steeply the velocity curve rises. A very steeply-rising curve could already reach an amplitude of $V_\mathrm{max}$ at the (relatively small) radius $R_\mathrm{fid}$ -- such a rotation curve would correspond to a point lying on the dashed line along $V_\mathrm{fid} = V_\mathrm{max}$ in the figure. A matter distribution with a cusp corresponds to a relatively steeply-rising circular velocity curve, such that by $R_\mathrm{fid}$ it is already approaching $V_\mathrm{max}$. For an \citetalias{NFW96} halo model appropriate for a dwarf galaxy with typical concentration \citep{Ludlow+16}, $V_\mathrm{c,fid}$ is typically about $0.65V_\mathrm{c,max}$; the precise relation is shown with a solid black line in the figure. A constant-density core in the matter distribution instead leads to a shallower, linear rise in rotation velocity within the core radius. Cores with sizes ${\gtrsim R_\mathrm{fid}}$ therefore have lower $V_\mathrm{fid}$ values. We note that $V_\mathrm{max}$ and $V_\mathrm{fid}$ are set by the shape of the total (dark plus baryonic) matter distribution. However, since the DM is often the dominant contributor to the gravitational force at all radii in dwarfs, in many cases these two parameters can in principle provide a direct measure of the DM density profile shape.

The galaxy LT-26 has a DM cusp; its steeply-rising circular velocity curve (Fig.~\ref{fig:rot_curves}, black curve in upper panel) reflects this. The values of $V_\mathrm{c,fid}$ and $V_\mathrm{c,max}$ measured from this curve are plotted in the left panel of Fig.~\ref{fig:loops} with a blue square symbol lying nearly directly on the line corresponding to an \citetalias{NFW96} halo model. The corresponding galaxy in the high SF density threshold simulation, HT-28, has had its DM cusp transformed into a core after repeated gas accretion and expulsion cycles. This is reflected in the more slowly-rising shape of its circular velocity curve (Fig.~\ref{fig:rot_curves}, black curve in lower panel). Since $V_\mathrm{c,max}$ is left essentially unchanged by the formation of the DM core, the slower rise of the rotation curve leads to a lower $V_\mathrm{c,fid}$ at approximately fixed $V_\mathrm{c,max}$, and the red square corresponding to the circular velocity curve of this galaxy lies below the black solid line in the left panel of Fig.~\ref{fig:loops}.

When these two galaxies are mock observed and modelled, the resulting rotation curves do not exactly recover the corresponding circular velocity curves -- in some cases, the discrepancy can be quite severe, as is visible in Fig.~\ref{fig:rot_curves}. In the left panel of Fig.~\ref{fig:loops}, each of the $24$ red and blue points corresponds to one of the $24$ rotation curves measured for LT-26 and HT-28, respectively. Consecutive observations (offset by $15\degr$ in $\Phi$) are linked by lines. The orientation labelled ${\Phi=0\degr}$, which we recall has no particular significance but is highlighted in Fig.~\ref{fig:rot_curves}, is larger than the other points and shown with error bars. The fractional uncertainty in both $V_\mathrm{\phi,max}$ and $V_\mathrm{\phi,fid}$ has no significant dependence on $\Phi$ and so is not shown for all points. For LT-26, the blue points lie systematically below the blue square symbol, i.e. $V_\mathrm{c,fid}$ is systematically underestimated by $V_\mathrm{\phi,fid}$ when modelling the mock observations, and there is considerable scatter across viewing angles in both $V_\mathrm{\phi,fid}$ and $V_\mathrm{\phi,max}$ -- a significantly greater scatter than the uncertainty associated with individual points. \citetalias{Oman+19} \citep[see also][]{Marasco+18} attributed the scatter to NCMs and the systematic underestimate of the thickness of the \ion{H}{i} discs of dwarf galaxies in the APOSTLE simulations. These conclusions also apply to the LT simulations, which are very similar to APOSTLE. The left panel of Fig.~\ref{fig:loops} illustrates that similar scattering effects are at play in our kinematic analyses of galaxies from the HT simulation however the underestimate of $V_\mathrm{c,fid}$ is not present, likely due to the thinner gas discs of HT systems (see Sec.~\ref{subsec:thick}).

\subsubsection{Rotation curve shape and central baryon dominance} \label{subsubsec:baryon_dom}

We also adopt two informative velocity ratios used by \citetalias{SantosSantos+20}: the rotation curve shape parameter, $\eta_{\mathrm{rot}}$, and the central baryonic matter fraction parameter, $\eta_{\mathrm{bar}}$.

$\eta_{\mathrm{rot}}$ condenses the information contained in the left panel of Fig.~\ref{fig:loops} into a single parameter, and therefore is a measure of how steeply a rotation curve rises. It is defined as:
\begin{equation} 
    \eta_{\mathrm{rot}} = \frac{V_{\mathrm{fid}}}{V_{\mathrm{max}}}.
    \label{eq:etarot}
\end{equation}
The circular velocity curve of an \citetalias{NFW96} DM halo has ${\eta_\mathrm{rot}\sim 0.65}$ (i.e. they lie near the solid black lines in the left and centre panels of Fig.~\ref{fig:loops}), while haloes hosting cores have correspondingly lower values.

The second velocity ratio we use is $\eta_{\mathrm{bar}}$, defined as:
\begin{equation} 
    \eta_{\mathrm{bar}} = \left(\frac{V_{\mathrm{b,fid}}}{V_{\mathrm{fid}}}\right)^{2},
    \label{eq:etabar}
\end{equation}
where ${V_{\mathrm{b,fid}} = V_{\mathrm{bar}}(R_{\mathrm{fid}})}$ is the contribution to the circular velocity due to baryonic matter (stars and gas) at $R_\mathrm{fid}$. Under the assumption of spherical symmetry, $\eta_\mathrm{bar}$ is equal to the baryon mass fraction within the fiducial radius. This assumption never holds exactly for late-type dwarfs, however, DM-dominated systems are likely reasonably close to spherical, such that $\eta_\mathrm{bar}$ serves as an approximate tracer of the importance of the baryons in setting the kinematics near the galactic centres. For simulated galaxies, we derive $V_\mathrm{b,fid}$ from the median gravitational acceleration due to baryon (star and gas) particles at a sample of points evenly distributed in azimuth in the disc midplane at $R_\mathrm{fid}$. For observed galaxies, it is derived from stellar (photometric) and \ion{H}{i} surface density profiles under the assumption of either razor-thin disc or vertically extended disc geometry \citep[for details, see][see also Appendix~E]{deBlok+08,Oh+15,Lelli+16}. Though it is possible to infer $V_{\mathrm{b,fid}}$ from our mock data cubes, with the possible addition of mock observations of the stellar component if it contributes a significant fraction of the mass, we do not attempt this in this work in order to focus on the role of the rotation curve measurement.

We plot these two velocity ratios against each other in the centre panel of Fig.~\ref{fig:loops}. As outlined above, points corresponding to steeply-rising rotation curves lie above those corresponding to slowly rising ones in this figure, and systems with a larger (apparent) contribution by baryons to the central mass content lie to the right of those centrally dominated by DM.

The values as determined from the circular velocity curves (and baryonic mass profiles) of LT-26 and HT-28 are shown with the blue and red squares, respectively. In addition to having a more slowly-rising circular velocity curve than LT-26 (as already seen in the left panel), HT-26 is also less centrally DM-dominated, with the baryons accounting for $\sim 40$~per~cent of the mass within $R_\mathrm{fid}$, compared to only $\sim 12$~per~cent in LT-26. As will be confirmed below, this is a feature typical of late-type dwarfs in the HT model: the higher gas density threshold for SF allows gas to condense in the galactic centres and become more tightly gravitationally bound, allowing a dense, massive gas disc to assemble\footnote{We note that this is generally true for both selected and excluded simulation samples, with galaxies excluded from kinematic analysis having a broadly similar diversity in $\eta_\mathrm{bar}$ to included galaxies from their respective model.} (see also Fig.~\ref{fig:properties}, where it can be seen that the HT dwarfs that we select for kinematic analysis are typically somewhat more gas-rich, and yet have smaller $R_\mathrm{HI}$, than their LT counterparts).

When $\eta_\mathrm{rot}$ and $\eta_\mathrm{bar}$ are plotted against each other for the ensemble of mock observations of a single galaxy, an anti-correlation emerges: rotation curves that rise more rapidly also give the appearance of a more DM-dominated central region. The reason for this is straightforward: if $V_\mathrm{c,fid}$ is underestimated by $V_\mathrm{\phi,fid}$, for example, the rotation curve rises more slowly, but the absolute contribution of the baryons to the circular velocity at the fiducial radius, $V_\mathrm{b,fid}$, remains fixed, so the galaxy appears more centrally baryon-dominated, and vice-versa for an overestimate of $V_\mathrm{c,fid}$. $V_\mathrm{b,fid}$ will remain fixed whether it is determined directly from simulation, as in this work, or from the \ion{H}{i} flux density and stellar photometry, as neither method is dynamical: $V_\mathrm{b,fid}$ is independent of $V_\mathrm{\phi,fid}$. Some scatter is introduced in the anti-correlation due to the fact that $V_\mathrm{c,max}$ may also be over- or underestimated by $V_\mathrm{\phi,max}$ during kinematic modelling,\footnote{We note that the changes in $V_\mathrm{\phi,max}$ also cause an adjustment to $V_\mathrm{\phi,fid}$ through the definition of $R_\mathrm{\phi,fid}$ (Eq.~\ref{eq:fiducial_things}).} represented by the error bars shown for the enlarged ${\Phi=0\degr}$ point.

The right panels of Fig.~\ref{fig:loops} illustrate that the values of $\eta_\mathrm{rot}$ and $\eta_\mathrm{bar}$ do not vary randomly as a function of the viewing angle $\Phi$: both oscillate in an approximately sinusoidal pattern with a period of $180\degr$. This demonstrates that the measurements of the rotation curves near the centres of these galaxies are affected by the presence of predominantly bisymmetric NCMs. Similar patterns are seen in all of the galaxies in our samples of LT and HT galaxies (see additional figures in Appendix~E). This is reminiscent of the analysis of APOSTLE galaxies by \citetalias{Oman+19} (see for instance their fig.~6) -- it is unsurprising that the very similar LT simulation which we use here exhibits the same behaviour. Apparently, galaxies in the HT model also have significant bisymmetric distortions of the gas flows near their centres.

\subsection{Rotation curve shape and baryon surface density} \label{subsec:surface_dens}

\begin{figure}
\begin{center}
\includegraphics[width=\columnwidth]{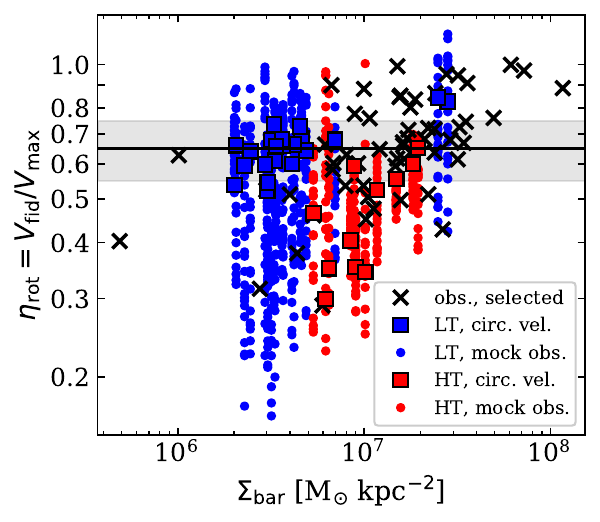}
\end{center}
\caption{Rotation curve shape parameter, $\eta_{\mathrm{rot}}$, as a function of effective baryon surface density, $\Sigma_{\mathrm{bar}}$. Measurements for selected galaxies from the LT and HT simulations are shown in blue and red, respectively. Measurements for selected observed galaxies from the THINGS and LITTLE~THINGS surveys are shown with black crosses. $\Sigma_{\mathrm{bar}}$ for simulated galaxies is calculated directly from the simulation (i.e. not from mock observations), and $\eta_{\mathrm{rot}}$ for the square markers is obtained from the circular velocity curves (derived from the gravitational acceleration field in the disc midplane). However, points show $\eta_{\mathrm{rot}}$ from each of the $24$ rotation curves obtained from mock observations of each galaxy. The solid black line shows the average $\eta_{\mathrm{rot}}$ expected for an \citetalias{NFW96} halo model and the grey band shows the $10^\mathrm{th}$--$90^\mathrm{th}$~percentile scatter around this value \citep{Ludlow+16}. With mock observation, LT and HT values align reasonably well with the weak positive correlation seen in the observed dwarfs.}
\label{fig:surface_density}
\end{figure}

Rotation curve shape is known to correlate with the surface brightness of galaxies (e.g. \citealt*{deBlok+96}; \citealt{Swaters+09,Swaters+12}; \citealt*{Lelli+13}), indicating that baryons may play a significant role in shaping their mass distributions. In Fig.~\ref{fig:surface_density} we show this correlation for both observed (black crosses) and simulated (blue and red markers) dwarf galaxies from our selected samples (see Sec.~\ref{subsec:Gal_selection}). We use $\eta_{\mathrm{rot}}$ to capture the shape of the circular velocity curve derived from the gravitational acceleration field in the disc midplane (squares) and from rotation curves from mock observations (points) as described in Sec.~\ref{subsec:rotcurshapes}. The `effective' baryon surface density, $\Sigma_{\mathrm{bar}}$, of each galaxy is ${\Sigma_{\mathrm{bar}} = M_{\mathrm{bar}} / 2\pi r_{\mathrm{b,half}}^{2}}$, where $M_{\mathrm{bar}}$ is the total baryonic mass and $r_{\mathrm{b,half}}$ is the radius of a sphere enclosing $0.5M_{\mathrm{bar}}$ -- both calculated directly from the distribution of particles in the simulations\footnote{Calculation of $\Sigma_{\mathrm{bar}}$ from mock $21$-cm observations alone is not possible. However, we find that $\Sigma_{\mathrm{HI}}$ derived from a mock observation is typically very close to the value obtained directly from the simulation particle distribution. We therefore have no reason to suppose that mock observation would significantly affect $\Sigma_{\star}$, and therefore the $\Sigma_{\mathrm{bar}}$ values shown in Fig.~\ref{fig:surface_density}.}.

In the observed population the correlation between $\eta_{\mathrm{rot}}$ and $\Sigma_{\mathrm{bar}}$ is primarily due to the fact that $\eta_{\mathrm{rot}}$ has a strong dependence on $V_\mathrm{max}$: in general, massive galaxies (i.e. high $V_\mathrm{max}$) are observed to have steeply-rising rotation curves \citepalias[large $\eta_\mathrm{rot}$; see e.g.][fig.~3]{SantosSantos+20}. Observed dwarf galaxies (${V_\mathrm{\phi,max} \lesssim 120/\,\mathrm{km}\,\mathrm{s}^{-1}}$), however, have a wide range of $\eta_{\mathrm{rot}}$ at fixed $\Sigma_{\mathrm{bar}}$. Baryon surface density therefore cannot be used to predict the presence of a core or cusp reliably in any particular dwarf galaxy. We consider now whether this picture changes when the simulated galaxies are `observed'.

HT galaxies within our interval of interest in $V_\mathrm{c,max}$ typically have intrinsically higher $\Sigma_{\mathrm{bar}}$ than LT galaxies (although there are two outliers), as expected due to their denser \ion{H}{i} discs. Both LT and HT galaxies, as measured using their `true' circular velocity curves, replicate the observed positive correlation between $\eta_{\mathrm{rot}}$ and $\Sigma_{\mathrm{bar}}$. However, neither replicate the observed scatter. The points in Fig.~\ref{fig:surface_density} show the approximate distribution of $\eta_{\mathrm{rot}}$ obtained for each galaxy when their rotation curves are instead measured from mock observations. Broadly speaking, these points better cover the observed range in $\eta_{\mathrm{rot}}$ and $\Sigma_\mathrm{bar}$, although it seems that the HT model struggles to reproduce galaxies observed to have high central baryon densities and very steeply rising rotation curves ($\eta_{\mathrm{rot}}\sim 1$).

\subsection{Which simulation better reproduces the observed galaxy population?} \label{subsec:errors_due_to_NCMs}

\begin{figure*}
\begin{center}
\includegraphics[width=0.935\textwidth]{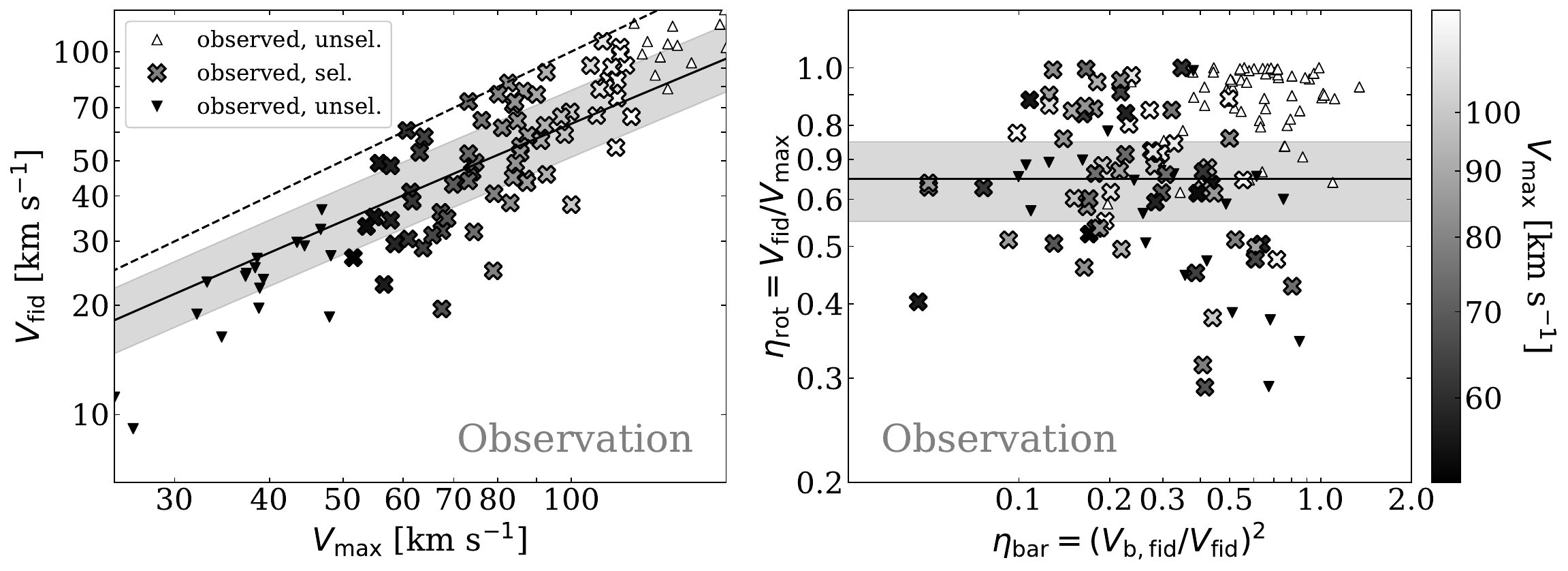}
\includegraphics[width=0.935\textwidth]{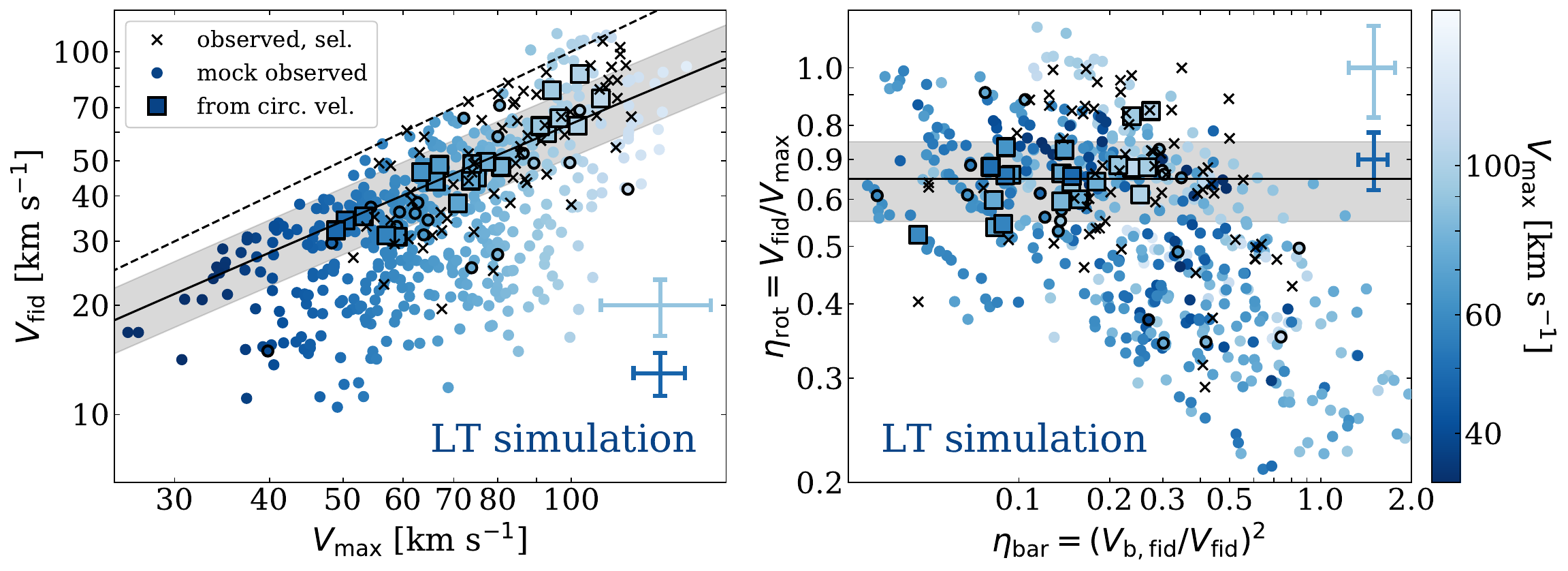}
\includegraphics[width=0.935\textwidth]{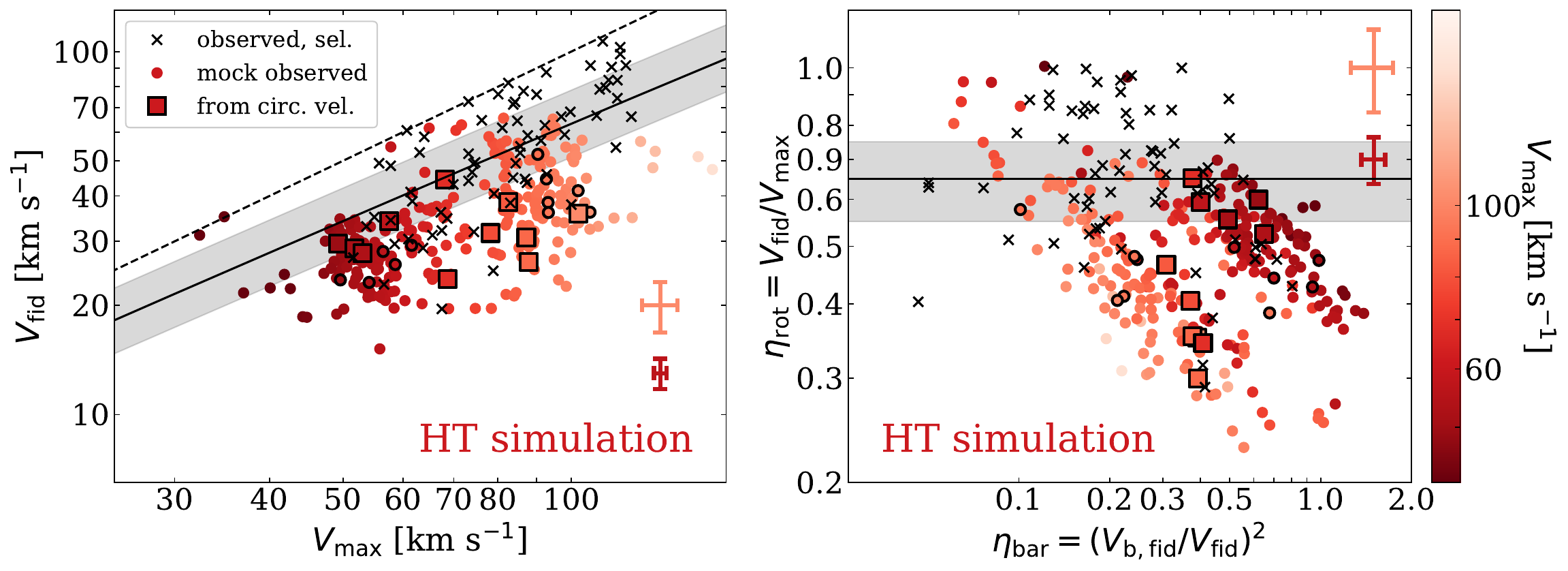}
\end{center}
\caption{\emph{Upper-left:} Inner `fiducial' velocity, $V_\mathrm{\phi,fid}$, (see Eq.~\ref{eq:fiducial_things} and Fig.~\ref{fig:rot_curves}) as a function of maximum velocity, $V_\mathrm{\phi,max}$, for a selection of galaxies observed in \ion{H}{i} with ${50<V_\mathrm{\phi,max}/\mathrm{km}\,\mathrm{s}^{-1}<120}$ shown as crosses, and unselected galaxies with $V_\mathrm{\phi,max}$ above and below this range shown as upward- and downward-pointed triangles, respectively (see Sec.~\ref{subsubsec:obs_sample}). Cross markers are coloured by $V_\mathrm{\phi,max}$ (redundant with the horizontal axis, but included for consistency with the upper-right panel). The dashed black line shows ${V_{\mathrm{fid}} = V_{\mathrm{max}}}$, while the solid black line shows the relation for an average \citetalias{NFW96} profile assuming the mass-concentration relation of \citet{Ludlow+16}, with $10^\mathrm{th}$--$90^\mathrm{th}$~percentile scatter shown as a grey band. \emph{Upper-right:} Rotation curve shape parameter, $\eta_{\mathrm{rot}}$, as a function of central baryon-to-total mass fraction $\eta_{\mathrm{bar}}$ (see Sec.~\ref{subsec:rotcurshapes}). Symbols and colours are as in the upper-left panel. The solid black line shows the average $\eta_{\mathrm{rot}}$ expected for \citetalias{NFW96} cusps and the grey band shows corresponding $10^\mathrm{th}$--$90^\mathrm{th}$~percentile scatter. \emph{Centre-left:} Similar to upper-left panel; lines and cross symbols (without colouration) are repeated. The square symbols are measured from the circular velocity curves ($V_\mathrm{c,max}$, $V_\mathrm{c,fid}$, derived from the gravitational acceleration field in the disc midplane) of the $21$ simulated LT galaxies selected for kinematic modelling. For each galaxy, there are $24$ points for the $24$ rotation curves ($V_\mathrm{\phi,max}$, $V_\mathrm{\phi,fid}$), one for each mock observation orientation, with points corresponding to ${\Phi=0\degr}$ given black borders. These give an impression of the range in rotation curve shapes which might be measured given the intrinsic shapes reflected by the square markers. Darker and lighter error bars show median fractional uncertainties (for ${\Phi=0\degr}$ mock observations) for galaxy subsamples with ${V_\mathrm{c,max}<75\,\mathrm{km}\,\mathrm{s}^{-1}}$ and ${V_\mathrm{c,max}\geq75\,\mathrm{km}\,\mathrm{s}^{-1}}$, respectively. \emph{Centre-right:} Similar to upper-right; lines and cross symbols (without colouration) are repeated. Blue squares and points are as in centre-left panel. \emph{Lower-left:} As centre-left, but for the $11$ selected, simulated HT galaxies. \emph{Lower-right:} As centre-right, but for selected HT galaxies.}
\label{fig:full}
\end{figure*}

In the upper panels of Fig.~\ref{fig:full} we show measurements from the full observational comparison sample in both the $V_\mathrm{\phi,fid}$--$V_\mathrm{\phi,max}$ (upper-left panel) and $\eta_\mathrm{bar}$--$\eta_\mathrm{rot}$ planes. The observed `selected' galaxies have the additional constraint ${50 < V_\mathrm{\phi,max} / \mathrm{km} \, \mathrm{s}^{-1} < 120}$ (the same as for the simulated galaxies) and are shown with crosses coloured by $V_\mathrm{\phi,max}$. These are repeated in the lower panels with small cross symbols. Galaxies with $V_\mathrm{\phi,max}$ above and below this range are shown with upward- and downward-pointed triangles, respectively, and are not repeated in the other panels.

Observed galaxies scatter widely around the solid black line marking the expectation for an \citetalias{NFW96} halo model (corresponding to ${\eta_\mathrm{rot}\approx 0.65}$). The scatter is significantly greater than that expected from the expected scatter in halo concentration (gray shaded band) -- this is the diversity of rotation curve shapes highlighted by \citet{Oman+15}.

Selected galaxies are also observed to have a diversity of central baryon mass fractions between ${\eta_\mathrm{bar}\approx 0.05}$~and~$0.7$ (upper-right panel). Observed dwarfs (darker symbols) exhibit a (weak) anti-correlation reminiscent of that seen in Fig.~\ref{fig:loops}, such that more centrally DM-dominated galaxies (i.e. lower $\eta_\mathrm{bar}$) have more steeply-rising rotation curves (i.e. higher $\eta_\mathrm{rot}$), and vice versa.  \citetalias{SantosSantos+20} note that this trend seems inconsistent, a priori, with BICC and SIDM models where almost all galaxies are predicted to have cores and cusps only form in baryon-dominated systems, and showed that mock observation alone could replicate this trend.

For any given observed galaxy, we are of course limited to a single sight line, such that an observational measurement leading to an equivalent of the points linked by lines in Fig.~\ref{fig:loops} is impossible. However, it is plausible that an ensemble of measurements for a collection of at least loosely similar galaxies (e.g. dwarfs) could preserve the trends outlined in Sec.~\ref{subsec:rotcurshapes}, independently of whether the galaxies have DM cusps, or cores produced by BICC.

In Fig.~\ref{fig:full} we show the $V_{\mathrm{c,max}}$ and $V_{\mathrm{c,fid}}$ values of the circular velocity curves (derived from the gravitational acceleration field in the disc midplane) of the 21~LT and 11~HT galaxies in our mock observed sample with the square symbols in the centre-left and lower-left panels, respectively. The measurements for the LT galaxies lie close to the solid black line marking the relation for an \citetalias{NFW96} halo of typical concentration: they all retain their central DM cusps (in the few galaxies with ${V_\mathrm{c,max}\gtrsim 90\,\mathrm{km}\,\mathrm{s}^{-1}}$, the haloes are contracted, driving up $V_\mathrm{c,fid}$). The measurements for the HT galaxies, on the other hand, lie systematically below the same line, reflecting the central DM deficit resulting from BICC.

If we assume that observational measurements (crosses) accurately trace the circular velocity curves of the observed galaxies, clearly neither the LT model nor the HT model is able to explain the observed diversity in terms of the circular velocity curves of the galaxies (squares). LT galaxies (blue squares) show much smaller scatter, and do not extend to values of $V_\mathrm{fid}$ as low as those of observed systems. HT galaxies (red squares), on the other hand, are all below the solid black line because of their pronounced cores. When the scatter induced by NCMs and the biases induced by disc thickness are taken into account, however, a different picture emerges.

We turn our attention to the distribution in $V_{\mathrm{\phi,max}}$ and $V_{\mathrm{\phi,fid}}$ of the rotation curves of the simulated galaxies (i.e. mock observed and modelled). The measurement corresponding to the orientation ${\Phi=0\degr}$ for each galaxy is shown with a point with black border in the centre-left and lower-left panels of Fig.~\ref{fig:full}. We recall that the direction defined by ${\Phi=0\degr}$ has no particular significance, so this is equivalent to choosing a random orientation (at fixed ${i=60\degr}$) for each galaxy. There are $24$ ($V_{\mathrm{\phi,max}}$, $V_{\mathrm{\phi,fid}}$) points for every square symbol, each corresponding to a different viewing angle, $\Phi$. This provides a reasonable estimate of the full range in the parameter space which can be reached by galaxies in the LT and HT models when they are `observed'. The scatter around the values derived from the circular velocity curves (squares) is substantial for both models. While the galaxies from the LT model essentially fully cover (and even somewhat exceed) the extent of the observed distribution, the HT galaxies only rarely reach the region at the highest $V_{\mathrm{\phi,max}}$ and $V_{\mathrm{\phi,fid}}$ sampled by the observations. The LT sample therefore seems to better align with observations in this space, suggesting that a model where every galaxy has a cusp, but appears at times to have a core, may be more plausible than a model where every galaxy has a core as large as those created in the HT model.

LT galaxies also reproduce the anti-correlation between $\eta_\mathrm{rot}$ and $\eta_\mathrm{bar}$ seen in the observed population of dwarfs quite naturally, spanning the full range covered by observations. The HT model, on the other hand, seems to struggle to reproduce observed dwarfs with steeply rising rotation curves (${\eta_{\mathrm{rot}} \sim 1}$) where the inner regions are DM-dominated (${\eta_{\mathrm{bar}} < 0.3}$). Furthermore, the HT galaxies with the highest $V_\mathrm{\phi,max}$ values in our sample preferentially lie toward lower $\eta_\mathrm{rot}$ and $\eta_\mathrm{bar}$, a feature absent for the LT galaxies, and running contrary to the observations: more massive galaxies typically have more steeply-rising rotation curves and are more centrally baryon-dominated \citep[e.g.][see also the upper-right panel of Fig.~\ref{fig:full}]{Broeils92,deBlok+08}. The discrepancy between the `observed' rotation curves of simulated galaxies and their circular velocity curves driven primarily by NCMs and thick \ion{H}{i} discs seems better able to reproduce the broad trends of the observed distribution of dwarfs in $\eta_\mathrm{rot}$ and $\eta_\mathrm{bar}$ than any of the other possible explanations considered by \citetalias[][]{SantosSantos+20} (i.e. SIDM, BICC without modelling errors, or the MDAR), particularly in the LT model where all galaxies have DM cusps.

In light of the results presented in this section, we argue that if the rotation curves of observed late-type dwarfs are significantly affected by errors due to NCMs and the geometric thickness of their \ion{H}{i} discs, the data favour a population of late-type dwarfs predominantly having DM cusps. In this interpretation, the observed DM `cores' in such objects are merely symptomatic of the failure of a rotation curve to trace the circular velocity curve of a galaxy. We note that we have not here explored the cores created by mechanisms other than BICC, such as SIDM.

\begin{figure*}
\centering
\includegraphics[width=\textwidth]{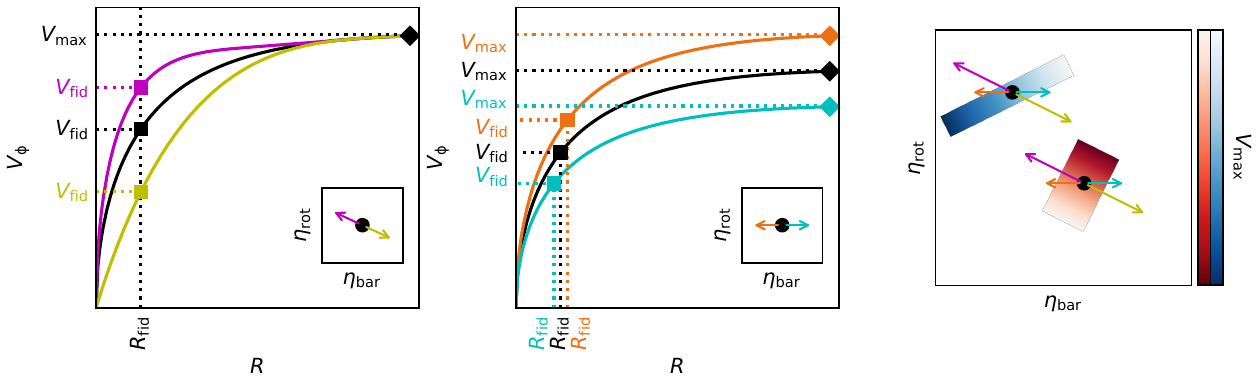}
\caption{A schematic representation of the connection between rotation curve shape and the parameters $\eta_\mathrm{rot}$ and $\eta_\mathrm{bar}$. \emph{Left:} An overestimate (magenta) or underestimate (yellow) of the inner rotation curve, such as might be induced by NCMs near the galactic centre. Since $V_\mathrm{\phi,max}$ is unaffected (and $V_\mathrm{b,fid}$ is assumed held fixed), the corresponding point in the $\eta_\mathrm{rot}$--$\eta_\mathrm{bar}$ plane is deflected along a line ${\eta_\mathrm{rot}\propto\eta_\mathrm{bar}^{-0.5}}$ (inset panel). \emph{Centre:} A global overestimate (orange) or underestimate (cyan) of the circular velocity curve, such as might be induced by a constant inclination error across the entire galactic disc. Since $V_\mathrm{\phi,fid}$ and $V_\mathrm{\phi,max}$ change approximately proportionally to one another, $\eta_\mathrm{rot}$ remains approximately constant, while $\eta_\mathrm{bar}$ is inversely proportional to $V_\mathrm{\phi,fid}^2$ (inset panel). \emph{Right:} The coloured patches are schematic representations of the distributions of square symbols in the centre-right and lower-right panels of Fig.~\ref{fig:full}: in the LT model, more massive dwarfs (increasing $V_\mathrm{c,max}$, paler blue colours) are increasingly centrally baryon-dominated (increasing $\eta_\mathrm{bar}$) and have increasingly steeply-rising rotation curves (increasing $\eta_\mathrm{rot}$). In the HT model the trend runs in approximately the opposite direction: more massive dwarfs are increasingly centrally DM-dominated and have more slowly-rising rotation curves. Deviations in rotation curve shape similar to those illustrated in the left panel drive the bulk of the scatter along the diagonal (magenta and yellow arrows, see also Fig.~\ref{fig:loops}). Deviations similar to those illustrated in the centre panel displace the corresponding point horizontally. For LT galaxies, this approximately opposes the intrinsic scalings with $V_\mathrm{c,max}$, e.g. an overestimate in $V_\mathrm{max}$ (${V_\mathrm{\phi,max}>V_\mathrm{c,max}}$; orange arrow) causes a shift towards a region occupied by galaxies with typically lower values of $V_\mathrm{c,max}$ (darker blue). For HT galaxies, however, an error in $V_\mathrm{\phi,max}$ instead exaggerates the intrinsic scalings with $V_\mathrm{c,max}$, explaining the origin of high-$V_\mathrm{\phi,max}$ mock observations lying at preferentially low-$\eta_\mathrm{rot}$ and low-$\eta_\mathrm{bar}$ in the HT model -- a feature absent in both the LT model and observations.}
\label{fig:schematic}
\end{figure*}

\section{Discussion}
\label{sec:Discussion}

In Sec.~\ref{subsec:errors_due_to_NCMs}, we argued that over- and underestimates of the central portions of rotation curves are common in our analysis of mock observations, and are the dominant driver of the scatter in the mock observed values of $\eta_\mathrm{rot}$ and $\eta_\mathrm{bar}$ along the direction of anti-correlation between $\eta_\mathrm{rot}$ and $\eta_\mathrm{bar}$. Discrepancies extending over the entire radial extent of the rotation curve are also relatively common, and lead to scatter in $\eta_\mathrm{bar}$ while leaving $\eta_\mathrm{rot}$ approximately constant. Overestimates of $\eta_\mathrm{bar}$ and/or underestimates of $\eta_\mathrm{rot}$ seem somewhat more common than the converse in our analysis; these can arise if the rotation curve is underestimated either in the centre, or at all radii. We also noted a preference for HT galaxies with higher $V_\mathrm{\phi,max}$ to lie at lower $\eta_\mathrm{rot}$ and lower $\eta_\mathrm{bar}$, a feature absent for LT galaxies. All of these trends and their connection to rotation curve shapes are summarised in Fig.~\ref{fig:schematic}. We next turn our attention to the possible physical origins of such effects, both for our simulated galaxies and their observed counterparts.

\subsection{Origins of discrepancies between rotation curves and circular velocity curves} \label{subsec:error_origins}

\subsubsection{Non-circular motions}
\label{subsubsec:disc_ncms}

If a model assuming pure circular rotation (as in our $^{\mathrm{3D}}${\sc barolo} analyses) is fit to a ring of gas including NCMs, each term in a Fourier series expansion of the NCMs will cause either an over- or underestimate of the rotation velocity, depending on the alignment between the phase angle and the line of sight. This implies that, provided the sight line is randomly chosen, there is no net preference for an error in the positive or negative direction, regardless of the values of the amplitudes $a_{m'}$ and $b_{m'}$ (see Eq.~\ref{eq:proj_ncm}).

As noted in our discussion of the right panels of Fig.~\ref{fig:loops} above, bisymmetric NCMs near the centres of our simulated galaxies are the driver of the bulk of the scatter in $\eta_\mathrm{rot}$ and $\eta_\mathrm{bar}$ in both the LT and HT models. These cannot, however, account for the fact that the values of these parameters derived from modelling mock observations lie preferentially below (for $\eta_\mathrm{rot}$) and above (for $\eta_\mathrm{bar}$) the corresponding `true' values obtained from the circular velocity curves of the galaxies.

\subsubsection{Disc thickness} \label{subsubsec:disc_disc_thickness}

The approach of most tilted-ring modelling algorithms, including $^{\mathrm{3D}}${\sc barolo}, is to constrain the parameters of each ring one at a time\footnote{The `regularisation' of the geometric parameters couples some parameters of adjacent rings, but the subsequent re-optimisation of the kinematic parameters still treats each ring independently of the others.}. When the model emission for a given ring is compared to the measured emission coming from the corresponding region on the sky, the assumption is that the emission comes from a fixed radial interval within the disc. For a razor-thin disc, the overlap between consecutive rings is minimal, even if a warp is present. For a geometrically thick disc, however, the emission in the region corresponding to a model ring includes mid-plane emission at the corresponding radius, but also emission coming from above and below the mid-plane, and therefore from other radii.

Attempting to make the model rings geometrically thick can further exacerbate the problem: adjacent rings will overlap in the plane of the sky and emission from a given location will be used to constrain multiple rings, even though the model implicitly assumes that the emission is to be decomposed into independent regions corresponding to each ring \citep[see e.g.][secs.~3.2, 7.1, for further discussion]{Iorio+17}. Ideally, all rings would be optimised simultaneously such that the overlap between geometrically thick rings could be accounted for, but the difficulty of parameter searches in high-dimensional spaces with multiple modes and strong degeneracies between parameters have so far precluded this.

\citet[][see their sec.~7.1]{Iorio+17} find that the net effects of $^{\mathrm{3D}}${\sc barolo}'s inability to correctly account for disc thickness during modelling are an underestimation of the rotation velocities at inner radii and an overestimation further out. This inner underestimation is clearly seen in the $V_{\mathrm{\phi,fid}}$ values of mock observed points being, on average, lower than the `true' squares in the left panels of Fig.~\ref{fig:full} -- especially in the LT sample, which has greater disc thicknesses (see Fig.~\ref{fig:disc_thicknesses}). Examples of the systematic overestimation of outer rotation velocities (i.e. ${V_{\mathrm{\phi,max}} > V_{\mathrm{c,max}}}$) can be seen in Fig.~\ref{fig:rot_curves} (blue and magenta lines above green and black at high $R$); this effect is not as strong as the former, however, it is still apparent in the left panels of Fig.~\ref{fig:full}, where points are, on average, shifted to the left of squares. The net underestimation of $V_{\mathrm{c,fid}}$ by $V_{\mathrm{\phi,fid}}$ and overestimation of $V_{\mathrm{c,max}}$ by $V_{\mathrm{\phi,max}}$ indicate that disc thickness effects are the main driver of the net underestimation of $\eta_{\mathrm{rot}}$ (and so overestimation of $\eta_{\mathrm{bar}}$, see Fig.~\ref{fig:schematic}) which, in turn, preferentially gives the appearance of DM cores.

\subsubsection{Inclination}
\label{subsubsec:incl}

An error in the overall inclination of a galaxy has a straightforward effect on the rotation curve, whose amplitude is inversely proportional to ${\sin i}$. An overestimate of $i$ therefore leads to both $V_\mathrm{\phi,max}$ and $V_\mathrm{\phi,fid}$ underestimating $V_\mathrm{c,max}$ and $V_\mathrm{c,fid}$. $\eta_\mathrm{rot}$ is nearly unchanged, except that the decrease in $V_\mathrm{\phi,max}$ reduces $R_\mathrm{\phi,fid}$, such that $V_\mathrm{c,fid}$ is underestimated by a slightly larger factor than $V_\mathrm{c,max}$ (see Fig.~\ref{fig:schematic}, centre panel), while $\eta_\mathrm{bar}$ is overestimated. Global inclination errors are therefore not the primary driver of the scatter of the measurements based on mock observations in Fig.~\ref{fig:full}.

The influence of radially-localised errors in inclination on correlations between $V_\mathrm{\phi,fid}$, $V_\mathrm{\phi,max}$, $\eta_\mathrm{rot}$, and $\eta_\mathrm{bar}$ are less easily predicted. However, `direct' misestimates of the local inclination in tilted ring modelling are relatively rare: tracing warps is, after all, precisely the task these models were designed to fulfill. Instead, local inclination errors most commonly arise in these models as a consequence of the models' inability to capture azimuthal asymmetries in the kinematics or, in other words, NCMs. Harmonic perturbations of order ${m=2}$, in particular, are strongly degenerate with the inclination angle \citep[e.g.][]{Schoenmakers+97}. An azimuthally-symmetric tilted-ring model therefore responds to an ${m=2}$ harmonic perturbation by adjusting the inclination angle to minimise the model residuals. If the perturbation has an amplitude and/or phase angle that vary with radius, the inclination error caused will also vary with radius. We prefer to think of such errors as being due to the non-circular nature of the gas orbits -- exactly of the sort described in Sec.~\ref{subsec:ncms} -- of which locally-incorrect inclination measurements are merely a symptom. A similar argument applies to degeneracies between the other geometric parameters (centre, systemic velocity, and position angle) and harmonic perturbations of other orders \citep[for an extensive discussion of such degeneracies, see][]{Schoenmakers+97}.

\subsubsection{Out-of-equilibrium kinematics}

That a rotation curve can be used to draw conclusions about the DM distribution within a galaxy is fundamentally predicated on the assumption that the motions of the kinematic tracers observed (in this context, \ion{H}{i} gas) are actually in dynamical equilibrium with the underlying gravitational potential. Simply put, if the gas is not orbiting at the local circular speed, a mass model based on a gas rotation curve loses its meaning. It is therefore important to know whether the \ion{H}{i} gas in the simulated galaxies actually orbits at the local circular speed.

We showed two examples in Fig.~\ref{fig:rot_curves}. We recall that the green curves show the median azimuthal velocity of \ion{H}{i}-bearing gas particles\footnote{The green curves are not corrected for pressure support, but the expected corrections are small, typically increasing $V_\mathrm{az}$ by only $5$~to~$10$~per~cent, nearly independently of radius \citep[e.g.][fig.~4]{Oman+19}. We reiterate that all mock observed rotation curves in this work, on which all our main conclusions rely, are corrected using the `asymmetric drift correction' computed by $^{\mathrm{3D}}${\sc barolo}.} as a function of radius for galaxies LT-26 and HT-28. In LT-26 this reaches, and follows for several kiloparsecs, the flat portion of the circular velocity curve, while in HT-28, it flattens near the edge of the \ion{H}{i} disc, which occurs before the circular velocity fully flattens. In both cases the media azimuthal velocity of the gas is close to the local circular velocity at $R_\mathrm{fid}$, but this is not the case for all galaxies in our sample. Similar figures for the full sample of galaxies selected for kinematic modelling are included in Appendix~E. Inspecting these by eye, we find that only ${\sim 7/21}$ of the sample drawn from the LT simulation have azimuthal velocity curves following the circular velocity curve at least as closely as the examples shown in Fig.~\ref{fig:rot_curves}, and we would describe only $2/21$ of them as following the circular velocity curve closely at all radii within the \ion{H}{i} disc. In the sample drawn from the HT simulation, the same fractions are somewhat higher: $10/11$ and $5/11$, respectively. However, we note that in both cases we have already removed from our sample those galaxies obviously unsuited to kinematic modelling following a visual inspection: $25$ from the LT simulation and $32$ from the HT simulation. These removed galaxies are preferentially ones where $V_\mathrm{az}$ is a poor tracer of $V_\mathrm{circ}$, and more galaxies are removed for the HT than for the LT model. The fraction of the total population of gas-rich, reasonably isolated dwarfs in these simulations where the \ion{H}{i} rotation curve is a good tracer of the circular velocity curve is therefore only about $\frac{1}{5}$ (in both simulations), and then only if the tilted ring model accurately recovers the azimuthal velocity of the \ion{H}{i} gas. In addition to the geometric thickness of the gas discs (see Sec.~\ref{subsubsec:disc_disc_thickness}), the tendency for $V_\mathrm{az}$ to underestimate $V_\mathrm{circ}$, especially in the inner regions, is a contributor to the systematic underestimates of $V_\mathrm{c,fid}$ reflected in Fig.~\ref{fig:full}.

Forming an impression of how many observed galaxies may have \ion{H}{i} azimuthal velocities differing significantly from their circular velocities is challenging, chiefly for two reasons. First, our impression is that the origins of such differences are diverse: mergers, interactions with companions, bulk flows driven by supernovae, gas accretion, an elongated potential due to a triaxial DM halo, disc instabilities, and interaction with the intergalactic medium can all disturb the ideal circular flow pattern. Ruling out each of these as a concern is labour intensive and realistically needs to be done on an object-by-object basis. Second, samples of observed galaxies selected for mass modelling are preferentially chosen to at least appear to be undisturbed and close to equilibrium. As outlined above, this mitigates how large a fraction of galaxies may be affected, but seems unlikely to reduce it to a negligible level. These issues are clearly of crucial importance in the interpretation of \ion{H}{i} rotation curve measurements; we plan to explore them further in future work \citep[see also][who find that BICC in their SMUGGLE galaxy formation model leads to $V_\mathrm{az}$ systematically underestimating $V_\mathrm{circ}$]{Jahn+21}.

\subsubsection{Imperfect model optimisation} \label{subsubsec:model_optimisation}

A source of error essentially unique to our analysis of mock observations, that is, not affecting the observed galaxies to which we compare them, is that we have largely allowed $^{\mathrm{3D}}${\sc barolo} to proceed automatically. That is, we do not revise and refine individual kinematic models to achieve the best possible description of the data in each case, as is customary in analyses of real galaxies. Our decision to largely automate the modelling process is driven primarily by necessity. We wish to consider of order $1000$ mock analyses to help bring out more subtle trends. For comparison, the few hundred rotation curves of highly-resolved galaxies in the literature have taken the field decades to produce. We note, however, that as the survey speeds of new telescopes continue to increase, so will the volume of available data to be modelled. Indeed, new surveys rely increasingly on automated kinematic modelling pipelines \citep[e.g.][]{Kamphuis+15,Ponomareva+21}.

We do not attempt to estimate directly how much such `automation errors' contribute to the overall scatter and/or systematic trends shown in Fig.~\ref{fig:full} \citep[but see][]{Kamphuis+15}. We focus instead on the physical drivers of the scatter and trends. As discussed above, NCM- and disc thickness-driven errors seem to be the dominant effects.

\subsection{Implications for other galaxy formation models in simulations} \label{subsec:gradient_generic}

\begin{figure*}
\centering
\includegraphics[width=\textwidth]{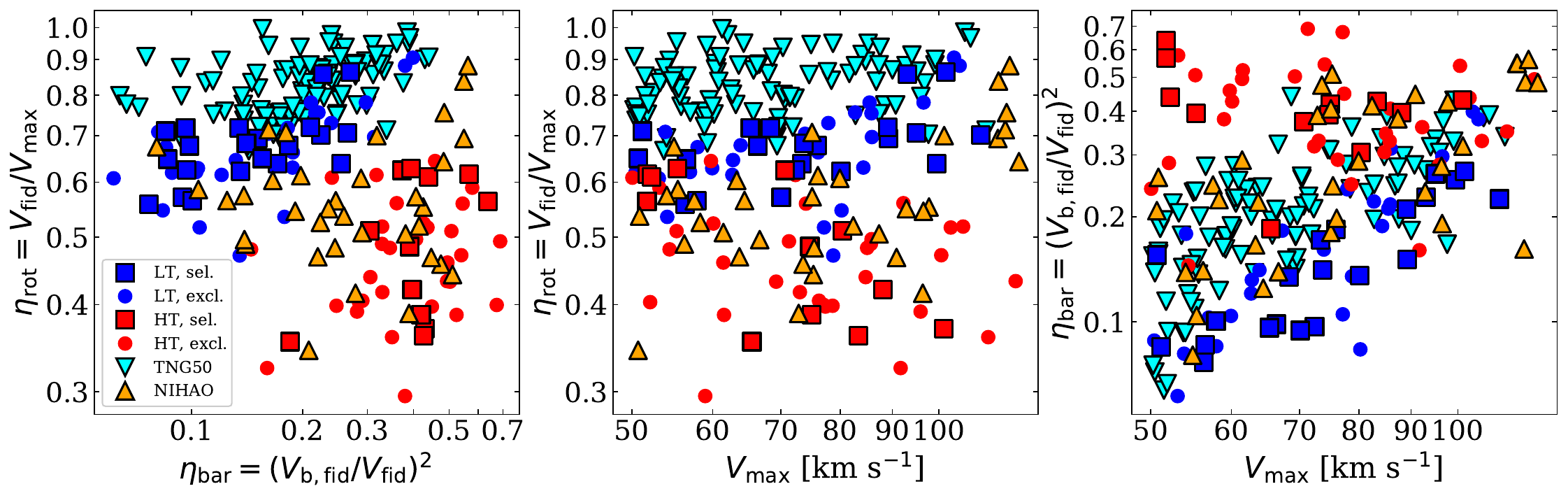}
\caption{\emph{Left panel:} Rotation curve shape parameter, $\eta_{\mathrm{rot}}$, as a function of central baryonic matter fraction, $\eta_\mathrm{bar}$, both determined from the (spherically-averaged) circular velocity curves. Galaxies from the LT and HT simulations selected for kinematic analysis are shown with blue and red square markers, respectively, while galaxies not selected are shown with points of the same colours. A selection of galaxies from the NIHAO simulation suite is shown with orange upward-pointed triangles, and a selection from the TNG50-1 simulation with cyan downward-pointed triangles. The LT and TNG50 populations occupy broadly similar regions of this parameter space, as do the HT and NIHAO populations. \emph{Centre panel:} $\eta_{\mathrm{rot}}$ as a function of maximum circular velocity, $V_{\mathrm{c,max}}$. Symbols are as in the left panel. LT and TNG50-1 galaxies have similarly low SF density thresholds, and all have steeply-rising rotation curves suggestive of DM cusps (${\eta_\mathrm{rot}\gtrsim 0.65}$). HT and NIHAO galaxies have similar, higher SF density thresholds, and often have slowly-rising rotation curves, suggestive of DM cores. \emph{Right panel:} $\eta_\mathrm{bar}$ against $V_{\mathrm{c,max}}$, both from circular velocity curves. Symbols are as in the other panels. The galaxy population in the LT simulation has a strong positive gradient in this space. The populations from both TNG and NIHAO have weaker positive gradients, while the population from the HT simulation has approximately constant $\eta_\mathrm{bar}$ across the range in $V_\mathrm{c,max}$ shown.}
\label{fig:etas-Vmax}
\end{figure*}

We next consider whether the trends for HT and LT galaxies illustrated in Fig.~\ref{fig:full} may be generic features of galaxy formation models where BICC does or does not occur in dwarfs, respectively, by comparing against two other models. For this, we use selections of galaxies from the IllustrisTNG and NIHAO simulation suites, without mock observation.

The TNG50-1 simulation \citep{Nelson+19,Pillepich+19} from the IllustrisTNG suite \citep{Nelson+18,Pillepich+18} is a high-resolution cosmological hydrodynamical simulation of a $(50\,\mathrm{cMpc})^3$ volume in a flat $\Lambda$CDM \citep{Planck14} cosmology, performed with the moving mesh code {\sc arepo} \citep{Springel10}. It has a baryon mass resolution of ${8.5 \times 10^{4} \,\mathrm{M}_{\sun}}$. The implementations of hydrodynamics, star formation, and supernova feedback are significantly different from those in the EAGLE model on which the LT and HT simulations are based. However, since SF in the TNG model occurs stochastically in regions exceeding a threshold density of ${n_{\mathrm{thr}} = 0.1\,\mathrm{cm}^{-3}}$, BICC does not operate efficiently, and in this sense the model is similar to the LT simulation. There are approximately $4500$ dwarf centrals with ${50 < V_{\mathrm{c,max}} / \mathrm{km}\,\mathrm{s}^{-1} < 120}$ and ${M_{\mathrm{HI}} > 10^{8}\,\mathrm{M_{\sun}}}$ in the TNG50 volume; we select $100$ of these at random as a representative sample. 

The NIHAO\footnote{Numerical Investigation of a Hundred Astrophysical Objects} project \citep{Wang+15} consists of a suite of $\sim100$ high-resolution cosmological zoom-in simulations in a flat $\Lambda$CDM \citet{Planck14} cosmology. The targets of the zoom regions are relatively isolated galaxies. The regions are evolved with a version of the ESF-{\sc gasoline}2 code (\citealt{Wang+15}; \citealt*{Wadsley+17}). NIHAO implements both supernova and early stellar feedback \citep[see][respectively]{Stinson+06,Stinson+13}. Star formation follows a Kennicutt-Schmidt SF law for ${T < 15000 \, \mathrm{K}}$ gas, with a density threshold of ${n_{\mathrm{thr}} = 10.3 \, \mathrm{cm}^{-3}}$. This is very close to the value in the HT simulation, and results in BICC in dwarf galaxies (\citealp{Dutton+16,SantosSantos+18,Dutton+19}; \citetalias{SantosSantos+20}). We select the $35$ galaxies from the NIHAO suite with ${50 < V_{\mathrm{c,max}} / \mathrm{km}\,\mathrm{s}^{-1} < 120}$.

In the left panel of Fig.~\ref{fig:etas-Vmax} we show $\eta_\mathrm{rot}$ against $\eta_\mathrm{bar}$ for LT, HT, TNG50, and NIHAO dwarfs, in all cases computed from their (spherically-averaged) circular velocity curves. The TNG50 galaxies invariably have steeply rising circular velocity curves ($\eta_\mathrm{rot} \gtrsim 0.7$), consistent with the presence of DM cusps. In the middle and right panels of the figure, we show $\eta_\mathrm{rot}$ and $\eta_\mathrm{bar}$, respectively, as a function of $V_\mathrm{c,max}$. This reveals that TNG50 dwarfs have more steeply rising rotation curves across the entire range in $V_\mathrm{max}$ considered. This may be due to their also having significantly higher central baryon fractions than LT galaxies across the same range in $V_\mathrm{c,max}$, which could contract their haloes. Given these qualitative similarities to the LT galaxies, we expect that if the TNG50 galaxies were mock observed, the $\eta_\mathrm{rot}$ and $\eta_\mathrm{bar}$ values obtained for their rotation curves would likely span a similar range as those in our sample of LT galaxies, offset to preferentially slightly higher $\eta_\mathrm{rot}$ and $\eta_\mathrm{bar}$. Our assessment is that this would likely result in covering the range spanned by the observed galaxies similarly to the galaxies from the LT model. It is therefore reasonable to speculate that our conclusions based on the LT model may be generalised to other qualitatively similar (in the sense that BICC does not occur) galaxy formation models.

Whereas the TNG50 and LT galaxies are broadly similar in the context of Fig.~\ref{fig:etas-Vmax}, comparing the NIHAO and HT galaxies in the same space reveals more pronounced qualitative differences between these two models. There are some similarities: in both models, BICC results in slowly rising rotation curves ($\eta_\mathrm{rot}\lesssim 0.65$), except in a few of the most massive NIHAO galaxies ($V_\mathrm{c,max}>100\,\mathrm{km}\,\mathrm{s}^{-1}$) -- this is likely related to their high central baryon fractions. However, whereas HT galaxies have slightly decreasing $\eta_\mathrm{rot}$ with increasing $V_\mathrm{c,max}$ and slightly decreasing $\eta_\mathrm{bar}$ with increasing $V_\mathrm{c,max}$, both of these trends are increasing for NIHAO galaxies. A consequence is that our finding that the most massive dwarfs in the HT model are preferentially found towards lower $\eta_\mathrm{rot}$ and $\eta_\mathrm{bar}$, which is difficult to reconcile with observations, does not apply to NIHAO galaxies. In addition, some NIHAO dwarfs occupy a region around $(\eta_\mathrm{bar},\eta_\mathrm{rot})\sim(0.15, 0.6)$, where no HT galaxies are found. We therefore speculate that if the NIHAO galaxies were observed, the resulting distribution would likely extend further to low $\eta_\mathrm{bar}$ and high $\eta_\mathrm{rot}$ than that for HT galaxies, and thus better span the range covered by observed galaxies\footnote{We note that the distribution of $\eta_\mathrm{rot}$ and $\eta_\mathrm{bar}$ derived from the circular velocity curves of NIHAO galaxies do not span the ranges observed, for instance near ${\eta_\mathrm{rot}\sim1.0}$ and ${\eta_\mathrm{bar}\sim0.15}$ \citepalias[see also][sec.~4.2.2]{SantosSantos+20}. Some additional source of scatter is therefore needed if this model is to explain the observations.}. In summary, it seems as though the conclusions which we draw based on the HT model cannot be generalised to all galaxy formation models where BICC occurs.

\subsubsection{Correlation between core size and stellar properties of galaxies}
\label{subsubsec:starcorrelations}

Other authors have highlighted that the size of cores created through BICC in simulations correlate with the stellar mass \citep{Governato+12}, stellar-to-halo mass ratio \citep{DiCintio+14} or star formation history \citep{Read+16}, with the largest cores found in simulated galaxies with $M_\star\sim10^8-10^9\,\mathrm{M}_\odot$, $M_\star/M_{200}\sim 10^{-2}$, and/or the most temporally extended star formation histories. If such correlations are clearly measured in observed galaxies, this will lend weight to interpretations involving BICC. While models parametrising these effects provide good descriptions of simulated galaxies with baryon-induced cores \citep[references above, and see also][]{Tollet+16,BenitezLlambay+19,Lazar+20}, and it is clear that they achieve a good description of some observed galaxies \citep[e.g.][]{Oh+15,Read+16,Read+19}, they so far fail to provide a comprehensive explanation for their overall scatter in central DM density slope, central rotation speed, or other measures of central DM content for galaxies in the regime where core formation peaks in these models \citep[][and see \citealp{Sales+22}, for a review]{SantosSantos+18,Relatores+19,SantosSantos+20,Frosst+22}. This leaves space for additional scatter due to effects such as those discussed in Sec.~\ref{subsec:error_origins}.

\section{Summary and conclusions}
\label{sec:Conclusion}

The diverse kinematics of observed local dwarf irregular galaxies is difficult to replicate directly in $\Lambda$CDM hydrodynamical simulations \citep[][]{Oman+15}. SIDM has been proposed as a possible explanation \citep[see e.g.][]{Kaplinghat+20}. More commonly, gravitational coupling of violent gas motions to the DM \citep[BICC;][]{NEF96,Read&Gilmore05,Pontzen&Governato12} have been invoked to explain the observed diversity \citep[e.g.][]{DiCintio+14,Brook15,Chan+15,Tollet+16,Read+16,PachecoArias21,Jahn+21}, with some concluding that this may not be sufficient \citep[e.g.][]{Bose+19,BenitezLlambay+19}. However, to our knowledge, no $\Lambda$CDM or SIDM galaxy formation model has yet been identified that can reproduce a realistic galaxy population including late-type dwarf galaxies with both the most steeply- and the most slowly-rising rotation curves observed. The observation that more centrally baryon-dominated late-type dwarfs seem to have more slowly-rising rotation curves (as quantified by the parameters $\eta_\mathrm{rot}$ and $\eta_\mathrm{bar}$), highlighted by \citetalias{SantosSantos+20}, proves to be especially challenging for BICC models to reproduce.

This discrepancy has been framed as a major challenge to the fiducial $\Lambda$CDM cosmology \citep[e.g.][]{Flores&Primack94,Moore94,Oman+15,Bullock&BoylanKolchin17,Sales+22}. However, recent studies have shown that systematic errors in the kinematic modelling of late-type dwarfs may be key to resolving this discrepancy (\citealt{Read+16,Iorio+17,Pineda+17}; \citetalias{Oman+19}; \citealt{Genina+20}). \citetalias{Oman+19} and \citetalias{SantosSantos+20} showed that discrepancies between the rotation curves and circular velocity curves of simulated late-type dwarfs induced primarily by the presence of non-circular flows in their \ion{H}{i} discs can plausibly reproduce the observed kinematic diversity, as well as the observed trend in $\eta_\mathrm{rot}$ and $\eta_\mathrm{bar}$, for galaxies with DM cusps drawn from the APOSTLE simulations. In the present study we build on this work by extending a similar analysis to many more simulated galaxies created using a galaxy formation model similar to APOSTLE (with low SF density threshold; LT), and a second model where BICC creates DM cores (with high SF density threshold; HT).

The two galaxy formation models are based on the EAGLE project \citep{Crain+15,Schaye+15}, with the gas density threshold for star formation modified to control the creation of DM cores \citep{BenitezLlambay+19}. We select galaxies with maximum circular velocities between ${50 \,\mathrm{and}\, 120 \, \mathrm{km}\,\mathrm{s}^{-1}}$, \ion{H}{i} masses exceeding ${10^{8} \, \mathrm{M}_{\sun}}$, and which are isolated from massive companions. By visual inspection, we remove those clearly disturbed which are unsuited to kinematic analysis, leaving a sample of galaxies with properties similar to those of observed late-type dwarfs. The simulated galaxies are mock observed from multiple viewing angles to produce \ion{H}{i} spectroscopic data cubes which are then analysed using a conventional 3D tilted-ring modelling tool in order to extract rotation curves.

We quantify rotation curve shapes by their amplitudes at outer radii ($V_\mathrm{\phi,max}$) and at an inner radius ($V_\mathrm{\phi,fid}$), and the ratio of the two ($\eta_\mathrm{rot}$). In addition, we quantify the dynamical importance of baryons near the galactic centre by the baryonic-to-total mass ratio in this region ($\eta_\mathrm{bar}$). Our main conclusions are summarised as:

\begin{enumerate}

\item The diversity in the rotation curve shapes as measured from the `true' circular velocity curves of galaxies (i.e. directly from the simulation outputs) from both the LT and the HT models falls far short of the observed diversity.

\item Assuming pure circular motion in galactic \ion{H}{i} discs in the presence of non-circular motions with realistic amplitudes introduces significant scatter in the recovered rotation curves for different viewing angles of the same galaxy. This scatter is similar in amplitude to the observed diversity in $V_\mathrm{\phi,fid}$ at fixed $V_\mathrm{\phi,max}$, and naturally leads to a correlation of the form ${\eta_\mathrm{rot}\propto\eta_\mathrm{bar}^{-0.5}}$. This scatter due to non-circular motions is symmetric around the true values of $\eta_\mathrm{rot}$ and $\eta_\mathrm{bar}$.

\item Assuming too thin an \ion{H}{i} disc typically causes an underestimation of the inner circular velocity curve (${V_\mathrm{\phi,fid}<V_\mathrm{c,fid}}$) and an overestimation of the outer circular velocity curve (${V_\mathrm{\phi,max}>V_\mathrm{c,max}}$), creating a bias towards slowly rising rotation curves (low $\eta_\mathrm{rot}$) and centrally high baryon fractions (high $\eta_\mathrm{bar}$). This bias is stronger in the LT galaxies, which have thicker \ion{H}{i} discs compared to the HT galaxies in our sample.

\item The combination of these two effects, which are dominant in our analysis of simulated galaxies, means that it is much more common for a galaxy with a DM cusp to appear to have a core than vice versa. It is likewise unusual, albeit not impossible, for simulated galaxies with cores to appear to have high central DM fractions. If the rotation curves of observed late-type dwarfs are similarly impacted -- which seems likely -- a late-type dwarf population where most galaxies have DM cusps is easier to reconcile with the available data than one where most galaxies have sizeable DM cores.

\item The mock observations of galaxies from the HT model with higher $V_\mathrm{\phi,max}$ lie preferentially at lower $\eta_\mathrm{rot}$ and/or lower $\eta_\mathrm{bar}$. No such trend is apparent in the LT model, or in observed late-type dwarfs. This lends some additional weight to a scenario where most late-type dwarfs host DM cusps.

\item An important caveat is that the HT simulation has significant relevant qualitative differences with respect to another galaxy formation model where BICC occurs (NIHAO) -- our conclusions based on this model are therefore unlikely to hold for all such models. The LT simulation, however, is broadly similar to another simulation where BICC does not occur (TNG50), suggesting that the conclusions based on this model may apply more generally to such models.

\end{enumerate}

Our results demonstrate that the possibility that the measured rotation curves of late-type dwarfs differ significantly from their circular velocity curves due to a combination of non-circular motions and geometrically thick \ion{H}{i} discs must be taken seriously. Indeed, these effects may plausibly be the dominant contributors to the scatter observed in the rotation curve shapes of these objects regardless of whether their DM haloes have central DM cusps, or cores. Our analysis shows that a scenario where all dwarf galaxies have central DM cusps remains viable; it may even be preferred.

\section*{Acknowledgements}

We thank the anonymous reviewer for detailed and constructive comments which helped to significantly improve this manuscript. We thank A.~Marasco for assistance with the calculation of harmonic amplitudes. We thank P.~Mancera~Pi{\~ n}a for kindly making outputs of $^{\mathrm{3D}}${\sc barolo} from analyses of nearby dwarfs publicly available. KAO, ISS and CSF acknowledge support by the European Research Council (ERC) through Advanced Investigator grant to C.S.~Frenk, DMIDAS (GA~786910). KAO acknowledges support by STFC through grant ST/T000244/1. ISS acknowledges support from the Arthur B. McDonald Canadian Astroparticle Physics Research Institute. ABL acknowledges European Research Council (ERC) under the European Union's Horizon 2020 research and innovation program (GA 101026328) and UNIMIB's Fondo di Ateneo Quota Competitiva (project 2020-ATESP-0133). This work used the DiRAC@Durham facility managed by the Institute for Computational Cosmology on behalf of the STFC DiRAC HPC Facility (www.dirac.ac.uk). The equipment was funded by BEIS capital funding via STFC capital grants ST/K00042X/1, ST/P002293/1, ST/R002371/1 and ST/S002502/1, Durham University and STFC operations grant ST/R000832/1. DiRAC is part of the National e\nobreakdash-Infrastructure. This research has made use of NASA's Astrophysics Data System.

\section*{Data availability}

The mock observed data cubes and numerical outputs of $^{\mathrm{3D}}${\sc barolo} for the galaxies listed in Supplementary~Tables~B1,~B2,~and~B3 and shown in Supplementary~Figs.~E1--E89, are available from the corresponding authors on reasonable request. Data used in this article from the following projects are publicly available: THINGS at \url{https://www2.mpia-hd.mpg.de/THINGS}, LITTLE~THINGS at \url{https://science.nrao.edu/science/surveys/littlethings}, SPARC at \url{http://astroweb.cwru.edu/SPARC}, and \citet{ManceraPina+21} at \url{https://pavel-mancera-pina.github.io}. Data from the IllustrisTNG project are available after registration at \url{https://www.tng-project.org}. Data provided by the NIHAO project were used with permission; for access, contact the authors of \citet{SantosSantos+18}. Software packages {\sc martini} and $^{\mathrm{3D}}${\sc barolo} are publicly available at \url{https://github.com/kyleaoman/martini} and \url{https://editeodoro.github.io/Bbarolo}, respectively.

\nocite{Swaters99}
\nocite{Wang+16}

\bibliographystyle{mnras}
\bibliography{bib1}

\appendix

\section*{Appendices}
\label{SecApp}

\noindent We include as supplementary materials, available from the publisher:

\smallskip

\textbf{Appendix A.} Comparisons with observed velocity dispersion profiles and surface density profiles.

\textbf{Appendix B.} Tables listing properties of galaxies selected for kinematic analysis, and detailing reasons some galaxies were excluded from this analysis.

\textbf{Appendix C.} A full listing of our $^{\mathrm{3D}}${\sc barolo} configuration.

\textbf{Appendix D.} A description of the algorithm used to define $V_\mathrm{\lowercase{\phi,max}}$ for rotation curves from simulation.

\textbf{Appendix E.} Summaries of the kinematic analysis of each individual galaxy, including figures similar to Figs.~\ref{fig:visualisations}, \ref{fig:rot_curves}, and \ref{fig:loops}.

\bsp
\label{lastpage}
\end{document}